\documentclass[twocolumn,times,tighten,twocolappendix]{aastex63}

\newcommand{\ergs}{$\rm erg~s^{-1}$}
\newcommand{\ergscm}{$\rm erg~s^{-1}~cm^{-2}$}
\newcommand{\ergscma}{$\rm erg~s^{-1}~cm^{-2}~\AA^{-1}$}
\newcommand{\fagn}{$F_{\rm AGN}$}
\newcommand{\fc}{$F_{\rm 5100}$}
\newcommand{\fcal}{$f_{\rm cal}$}
\newcommand{\fgal}{$F_{\rm gal}$}
\newcommand{\fgalabs}{$F_{\rm gal,abs}$}
\newcommand{\fph}{$F_{\rm phot}$}
\newcommand{\fhb}{$F_{\rm H\beta}$}
\newcommand{\feii}{Fe {\sc ii}}
\newcommand{\fvar}{$F_{\rm var}$}
\newcommand{\ha}{H$\alpha$}
\newcommand{\heii}{He {\sc ii}}
\newcommand{\hb}{H$\beta$}
\newcommand{\kms}{$\rm km~s^{-1}$}
\newcommand{\mbh}{$M_{\rm BH}$}
\newcommand{\oiii}{[O {\sc iii}]}
\newcommand{\rmax}{$r_{\rm max}$}
\newcommand{\sigc}{$\sigma_{\rm 5100}$}
\newcommand{\sigcal}{$\sigma_{\rm cal}$}
\newcommand{\sighb}{$\sigma_{\rm H\beta}$}
\newcommand{\sline}{$\sigma_{\rm line}$}
\newcommand{\tph}{$\tau_{\rm ph}$}
\newcommand{\tsp}{$\tau_{\rm sp}$}

\shorttitle{Reverberation for 15 PG Quasars}
\shortauthors{Hu et al.}

\begin{document}

\title{Supermassive Black Holes with High Accretion Rates in Active Galactic
Nuclei. XII. \\ Reverberation Mapping Results for 15 PG Quasars from a
Long-Duration High-Cadence Campaign}

\correspondingauthor{Jian-Min Wang}
\email{wangjm@ihep.ac.cn}

\author{Chen Hu}
\affil{Key Laboratory for Particle Astrophysics, Institute of High Energy
Physics, Chinese Academy of Sciences, 19B Yuquan Road, Beijing 100049, China}

\author{Sha-Sha Li}
\affil{Key Laboratory for Particle Astrophysics, Institute of High Energy
Physics, Chinese Academy of Sciences, 19B Yuquan Road, Beijing 100049, China}
\affil{School of Astronomy and Space Science, University of Chinese Academy of
Sciences, 19A Yuquan Road, Beijing 100049, China}

\author{Sen Yang}
\affil{Key Laboratory for Particle Astrophysics, Institute of High Energy
Physics, Chinese Academy of Sciences, 19B Yuquan Road, Beijing 100049, China}
\affil{School of Astronomy and Space Science, University of Chinese Academy of
Sciences, 19A Yuquan Road, Beijing 100049, China}

\author{Zi-Xu Yang}
\affil{Key Laboratory for Particle Astrophysics, Institute of High Energy
Physics, Chinese Academy of Sciences, 19B Yuquan Road, Beijing 100049, China}
\affil{School of Astronomy and Space Science, University of Chinese Academy of
Sciences, 19A Yuquan Road, Beijing 100049, China}

\author{Wei-Jian Guo}
\affil{Key Laboratory for Particle Astrophysics, Institute of High Energy
Physics, Chinese Academy of Sciences, 19B Yuquan Road, Beijing 100049, China}
\affil{School of Astronomy and Space Science, University of Chinese Academy of
Sciences, 19A Yuquan Road, Beijing 100049, China}

\author{Dong-Wei Bao}
\affil{Key Laboratory for Particle Astrophysics, Institute of High Energy
Physics, Chinese Academy of Sciences, 19B Yuquan Road, Beijing 100049, China}
\affil{School of Astronomy and Space Science, University of Chinese Academy of
Sciences, 19A Yuquan Road, Beijing 100049, China}

\author{Bo-Wei Jiang}
\affil{Key Laboratory for Particle Astrophysics, Institute of High Energy
Physics, Chinese Academy of Sciences, 19B Yuquan Road, Beijing 100049, China}
\affil{School of Astronomy and Space Science, University of Chinese Academy of
Sciences, 19A Yuquan Road, Beijing 100049, China}

\author{Pu Du}
\affil{Key Laboratory for Particle Astrophysics, Institute of High Energy
Physics, Chinese Academy of Sciences, 19B Yuquan Road, Beijing 100049, China}

\author{Yan-Rong Li}
\affil{Key Laboratory for Particle Astrophysics, Institute of High Energy
Physics, Chinese Academy of Sciences, 19B Yuquan Road, Beijing 100049, China}

\author{Ming Xiao}
\affil{Key Laboratory for Particle Astrophysics, Institute of High Energy
Physics, Chinese Academy of Sciences, 19B Yuquan Road, Beijing 100049, China}

\author{Yu-Yang Songsheng}
\affil{Key Laboratory for Particle Astrophysics, Institute of High Energy
Physics, Chinese Academy of Sciences, 19B Yuquan Road, Beijing 100049, China}
\affil{School of Astronomy and Space Science, University of Chinese Academy of
Sciences, 19A Yuquan Road, Beijing 100049, China}

\author{Zhe Yu}
\affil{Key Laboratory for Particle Astrophysics, Institute of High Energy
Physics, Chinese Academy of Sciences, 19B Yuquan Road, Beijing 100049, China}
\affil{School of Astronomy and Space Science, University of Chinese Academy of
Sciences, 19A Yuquan Road, Beijing 100049, China}

\author{Jin-Ming Bai}
\affil{Yunnan Observatories, The Chinese Academy of Sciences, Kunming 650011,
China}

\author{Luis C. Ho}
\affil{Kavli Institute for Astronomy and Astrophysics, Peking University,
Beijing 100871, China}
\affil{Department of Astronomy, School of Physics, Peking University, Beijing
100871, China}

\author{Michael S. Brotherton}
\affil{Department of Physics and Astronomy, University of Wyoming, Laramie, WY
82071, USA}

\author{Jes\'us Aceituno}
\affil{Centro Astronomico Hispano Alem\'an, Sierra de los filabres sn, E-04550
Gergal, Almer\'ia, Spain}
\affil{Instituto de Astrof\'isica de Andaluc\'ia (CSIC), Glorieta de la
astronom\'ia sn, E-18008 Granada, Spain}

\author{Hartmut Winkler}
\affil{Department of Physics, University of Johannesburg, PO Box 524, 2006
Auckland Park, South Africa}

\author{Jian-Min Wang}
\affil{Key Laboratory for Particle Astrophysics, Institute of High Energy
Physics, Chinese Academy of Sciences, 19B Yuquan Road, Beijing 100049, China}
\affil{School of Astronomy and Space Science, University of Chinese Academy of
Sciences, 19A Yuquan Road, Beijing 100049, China}
\affil{National Astronomical Observatories of China, The Chinese Academy of
Sciences, 20A Datun Road, Beijing 100020, China}

\collaboration{18}{(SEAMBH collaboration)}

\begin{abstract}

  We present the first results from long-term high-cadence spectroscopic
  monitoring of 15 PG quasars with relatively strong \feii\ emission as a part
  of a broader reverberation mapping campaign performed with the Calar Alto
  Observatory 2.2m telescope. The $V$-band, 5100 \AA\ continuum, and \hb\
  broad emission line light curves were measured for a set of quasars for
  between dozens to more than a hundred epochs from May 2017 to July 2020.
  Accurate time lags between the variations of the \hb\ broad line fluxes and
  the optical continuum strength are obtained for all 15 quasars, ranging from
  $17.0_{-3.2}^{+2.5}$ to $95.9_{-23.9}^{+7.1}$ days in the rest frame. The
  virial masses of the central supermassive black holes are derived for all 15
  quasars, ranging between $0.50_{-0.19}^{+0.18}$ and $19.17_{-2.73}^{+2.98}$
  in units of $10^7 M_\odot$. For 11 of the objects in our sample, this is the
  first reverberation analysis published. Of the rest, two objects have been
  the subject of previous reverberation studies, but we determine time lags
  for these that are only half as long as found in the earlier investigations,
  which had only been able to sample much more sparsely. The remaining two
  objects have previously been monitored with high sampling rates. Our results
  here are consistent with the earlier findings in the sense that the time lag
  and the line width vary inversely consistent with virialization. 

\end{abstract}


\section{Introduction}

Reverberation mapping \citep{blandford82} is a widely applied technique
enabling the determination of the masses of the supermassive black holes in
active galactic nuclei (AGNs) by observing the reverberation between the
variations of the broad emission lines and the continuum. To date the time lag
of \hb\ emission lines has been confirmed and measured in only $\sim$100 AGNs
(see \citealt{du19} for a compilation), as the process of monitoring the
spectrum with sufficient time sampling over lengthy periods is very
telescope-time-consuming. It has been highlighted in several studies that the
broad-line region (BLR) radius ($R_{\rm BLR}$) calculated from the measured
time lag can substantially deviate from the true value if the light curves do
not span long enough periods (see, e.g., the simulations in \citealt{goad14}),
or if the time between successive observations is too large (e.g.,
\citealt{grier08,hu20b,hu20a}).

There are very few previous reverberation mapping data sets that meet both the
requirements of a lengthy observation span and high cadence. For example, the
campaign of \citet{kaspi00} lasted $\sim$7.5 years, which is more than
sufficient to meet the first requirement, but the sampling intervals are of
the order of a few tens of days \citep{chelouche14} which is far longer than
optimum cadence. As highlighted by \citet{grier12}, the current database of
AGN nuclear parameters determined through reverberation mapping requires both
checking and revision through both new and sampling-improved measurements.

The SEAMBH collaboration (super-Eddington accreting massive black hole;
\citealt{du14}) started a long-term reverberation mapping campaign in May 2017
at the Calar Alto Observatory in Spain. Dozens of targets, mainly PG
(Palomar-Green Survey; \citealt{schmidt83}) quasars, have been monitored using
the Centro Astron\'omico Hispano-Alem\'an (CAHA) 2.2m telescope. The first
outcomes of this campaign, which presented the analysis of the data for PG
2130+099 and PG 0026+129, have been presented in \citet{hu20b} and
\citet{hu20a} respectively. The high cadence and long time sequence not only
enabled the measurement of more reliable time lags than previously reported,
but also revealed unexpected physical insights, including the fast change of
the BLR structure in PG 2130+099 \citep{hu20b} and the two distinct BLRs in PG
0026+129 \citep{hu20a}. By July 2020 sufficient data had been secured for
reliable measurements of \hb\ time lags in another 15 PG quasars. In 11 of
these quasars, \hb\ time lags are presented for the first time.

This paper is the first planned for presenting the study of these 15 PG
quasars. It focuses on the observations (Section \ref{sec-obs}), light curves
of the continuum and \hb\ (Section \ref{sec-lc}), \hb\ time lags given by
basic time series analysis (Section \ref{sec-ccf}), and the estimated black
hole masses (Section \ref{sec-mass}) of these objects. Section \ref{sec-cmp}
compares the results determined here for four objects for which previous
reverberation mapping campaigns had been carried out with the corresponding
outcomes from these earlier studies. Further analysis on the data of our
sample of these 15 PG quasars, including light curves and time lags of
emission lines other than \hb\ by detailed spectral fitting, host-galaxy
contamination using \textit{Hubble Space Telescope (HST)} images,
velocity-resolved delays and possible velocity-delay maps, will be presented
in future contributions.

\section{Observations and Data Reduction}
\label{sec-obs}

\subsection{Sample}

Our targets are part of the list of PG quasars studied by \citet{boroson92}.
The dimensionless accretion rates were calculated for all the 87 objects in
that list using equation (2) from \citet{du15} with the continuum luminosities
and line widths measured from the single-epoch spectra (the traditional BLR
radius--luminosity relation of \citealt{bentz13} was used to estimate the
radius). Then we selected our sample for reverberation mapping based on their
dimensionless accretion rates (in a sequence of high to low) and their
observability. At this stage we are monitoring about 40 PG quasars,
corresponding to the half of the \citet{boroson92} sample with the highest
accretion rates. All targets in our sample show relatively strong \feii\ and
comparatively weak \oiii, which are considered a spectral signature typical of
AGN with high accretion rates \citep[e.g.,][]{boroson02,shen14}.

\begin{deluxetable*}{llllccrcr@{~--~}l}
  \tablewidth{0pt}
  \tablecaption{Objects and Observations
  \label{tab-obs}}
  \tablehead{
  \colhead{Object} & \colhead{Other Name} &
  \colhead{$\alpha_{2000}$} & \colhead{$\delta_{2000}$} &
  \colhead{$z$} & \colhead{$A_V$} & 
  \colhead{$N_{\rm obs}$} & \colhead{$T_{\rm median}$} &
  \multicolumn{2}{c}{JD}
  \\
  \colhead{} & \colhead{} & \colhead{} & \colhead{} & \colhead{} &
  \colhead{(mag)} & \colhead{} & \colhead{(days)} &
  \multicolumn{2}{c}{(-245000)}
  \\
  \colhead{(1)} & \colhead{(2)} & \colhead{(3)} & \colhead{(4)} &
  \colhead{(5)} & \colhead{(6)} & \colhead{(7)} & \colhead{(8)} &
  \multicolumn{2}{c}{(9)}
  } 
  \startdata
PG 0003+199 &    Mrk 335 & 00h06m19.52s & +20d12m10.5s & 0.0259 & 0.096 &   44 &  3 & 8693 & 8893 \\
PG 0804+761 &            & 08h10m58.60s & +76d02m42.5s & 0.1005 & 0.097 &  149 &  5 & 8017 & 9036 \\
PG 0838+770 & VII Zw 244 & 08h44m45.31s & +76d53m09.7s & 0.1316 & 0.077 &  102 &  6 & 8016 & 8992 \\
PG 1115+407 &            & 11h18m30.27s & +40d25m54.0s & 0.1542 & 0.045 &   62 &  6 & 8453 & 9032 \\
PG 1322+659 &            & 13h23m49.52s & +65d41m48.2s & 0.1678 & 0.053 &   87 &  9 & 7883 & 8857 \\
PG 1402+261 &    TON 182 & 14h05m16.21s & +25d55m34.1s & 0.1643 & 0.043 &   41 &  7 & 7889 & 8327 \\
PG 1404+226 &            & 14h06m21.88s & +22d23m46.2s & 0.0972 & 0.061 &   92 &  5 & 7916 & 9036 \\
PG 1415+451 &            & 14h17m00.83s & +44d56m06.4s & 0.1132 & 0.024 &   93 &  7 & 7911 & 9041 \\
PG 1440+356 &    Mrk 478 & 14h42m07.46s & +35d26m22.9s & 0.0770 & 0.039 &  109 &  6 & 7883 & 9040 \\
PG 1448+273 &            & 14h51m08.76s & +27d09m26.9s & 0.0646 & 0.080 &   94 &  6 & 7887 & 8862 \\
PG 1519+226 &            & 15h21m14.26s & +22d27m43.9s & 0.1351 & 0.118 &   79 &  7 & 7886 & 8757 \\
PG 1535+547 &    Mrk 486 & 15h36m38.36s & +54d33m33.2s & 0.0385 & 0.040 &   36 &  4 & 7883 & 8077 \\
PG 1552+085 &            & 15h54m44.58s & +08d22m21.5s & 0.1187 & 0.113 &   66 &  7 & 7882 & 8760 \\
PG 1617+175 &    Mrk 877 & 16h20m11.29s & +17d24m27.7s & 0.1144 & 0.114 &   34 &  5 & 7885 & 8057 \\
PG 1626+554 &            & 16h27m56.12s & +55d22m31.5s & 0.1316 & 0.016 &   67 &  5 & 8536 & 9043
  \enddata
  \tablecomments{
  Columns (1) and (2) give the PG designation and, where applicable, a common
  alternative name for each object. Columns (3) and (4) list the right
  ascension and the declination. Column (5) lists the redshift determined by
  the narrow \oiii\ lines in our spectra.  Column (6) lists the Galactic
  extinction in the $V$ band from \citet{schlafly11}.  The number of observing
  epochs, median time between spectra, and time coverage of our spectroscopic
  observations are given for each object in columns (7) to (9).
  }
\end{deluxetable*}

Up to July 2020, reliable BLR reverberation lags had been secured for 17
objects. The results for PG 2130+099 and PG 0026+129 were presented in
\citet{hu20b} and \citet{hu20a}, respectively. The other 15 objects, covered
by this paper, are listed in Table \ref{tab-obs} together with a summary of
some of the observational details. Note that the redshifts $z$ (Column 5) are
defined by the narrow \oiii\ $\lambda\lambda$4959,5007 lines in the mean
spectrum of each target. The $V$-band Galactic extinctions $A_V$ listed in
Column (6) are from the NASA/IPAC Extragalactic Database%
\footnote{\url{https://ned.ipac.caltech.edu/}}
determined by \citet{schlafly11}. The majority of our targets have
declinations between 13h and 15h, and are among the first dozen objects
observed when the campaign began in May 2017. They were observed for roughly
100 epochs extending over more than 1000 days (three years).

The details of the observation strategy and data reduction have been reported
in detail in \citet{hu20b} and also in \citet{hu20a}, so we only present a
summary thereof below.

\subsection{Spectroscopy}
\label{sec-spec}

The spectra were taken at the CAHA 2.2m telescope using the Calar Alto Faint
Object Spectrograph (CAFOS) with Grism G-200, using a long slit set at a width
of 3$\farcs$0. The slit was rotated to an oriention allowing the simultaneous
observation of the quasar and a nearby comparison star with comparable
brightness, which was used to enable the flux calibration. The data were
reduced following standard procedures, including bias-removal, flat-field
corrections using dome flats and wavelength calibration by HgCd/He/Rb lamps.
The spectra of both the quasar and the comparison star are extracted in fixed
apertures of 10$\farcs$6$\times$3$\farcs$0, covering the wavelength range of
4000--8000 \AA\ with a dispersion of 4.47 \AA\ pixel$^{-1}$.

The flux calibration of the quasar spectrum was performed by fitting a
sensitivity function determined from the comparison star spectrum for each
exposure (See \citealt{hu20b} for the details of generating the fiducial
spectrum of the comparison star). Utilizing this procedure, the flux can be
calibrated to an accuracy better than $\sim$3\% for most of the observed
spectral range, as estimated in \citet{hu20b,hu20a}. The accuracy of our flux
calibration is also demonstrated by the consistency between the two light
curves of the spectroscopic 5100 \AA\ continuum and the photometric $V$-band
continuum, which were calibrated independently (see Section \ref{sec-lc}
below).

The actual spectral resolution is indicated by whichever is smaller of the
3$\farcs$0 slit width and the seeing during the exposure. The seeing was less
than 3$\farcs$0 most of the time. Seeing can be swiftly variable on some
nights at Calar Alto. Thus, the spectral resolution may change in spectra
taken at different epochs. On average, the instrumental broadening is
$\sim$1000 \kms\ for full width at half-maximum (FWHM), determined by
comparing the widths of narrow \oiii\ $\lambda$5007 lines in our spectra for
several objects with those obtained from high-spectral-resolution spectroscopy
\citep{hu20b}. The broadening changes in different spectra are evident in
the varying widths of the narrow emission lines, and may influence the
root-mean-square residual (rms) spectra to have narrower \hb\ profiles. Thus,
we measured the velocity width and shift of \oiii\ lines for each
individual-epoch spectrum [by the simple spectral fitting used for
recalibration in \citet{hu16}], and then convolved the flux-calibrated
spectrum with a corresponding Gaussian to unify shifting and broadening.

\begin{figure*}
  \centering
  \includegraphics[width=0.9\textwidth]{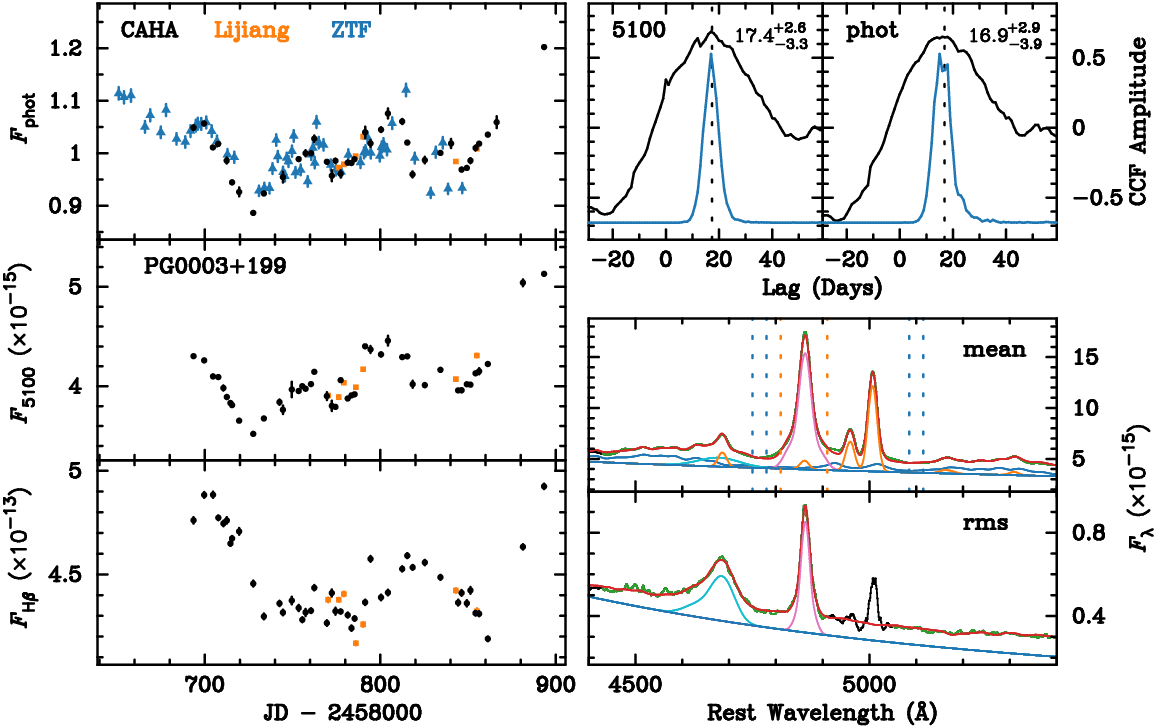}
  \caption{
  PG 0003+199 light curves, CCFs, mean and rms spectra. Left column, from top
  to bottom: light curves of the photometric continuum flux (\fph) in
  arbitrary linear units, the spectroscopic 5100 \AA\ continuum flux (\fc) in
  units of $10^{-15}$ \ergscma, and the integrated \hb\ flux (\fhb) in units
  of $10^{-13}$ \ergscm. The solid black circles, orange squares, and blue
  triangles denote the observations from CAHA, Lijiang, and ZTF, respectively.
  Right column, top panels: CCFs (in black) and corresponding CCCDs (in blue)
  for \fhb\ with respect to \fc\ and \fph\, respectively. The time lags (in
  the observed frame) and their errors are displayed in the panels, and marked
  by the vertical dotted lines. Right column, bottom panels: mean and rms
  spectra (green) in the rest-frame wavelength after the Galactic extinction
  correction, and decomposition by spectral fitting. The best-fit model (red)
  is composed by the AGN power law (blue), \feii\ (blue), broad \hb\
  (magenta), narrow lines (orange), broad \heii\ (cyan), and the host galaxy
  (low and out of the panel window, its strength can be seen by the difference
  between the fluxes of the best-fit model and the sum of AGN continuum and
  \feii\ at 5100 \AA). The region around the \oiii\ lines in the rms spectrum
  (pixels in black) is excluded in the fits, see the text in Section
  \ref{sec-mass} for details. The vertical dotted lines mark the continuum and
  \hb\ integration windows.
  }
  \label{fig-lc0003}
\end{figure*}

\begin{figure*}
  \centering
  \includegraphics[width=0.9\textwidth]{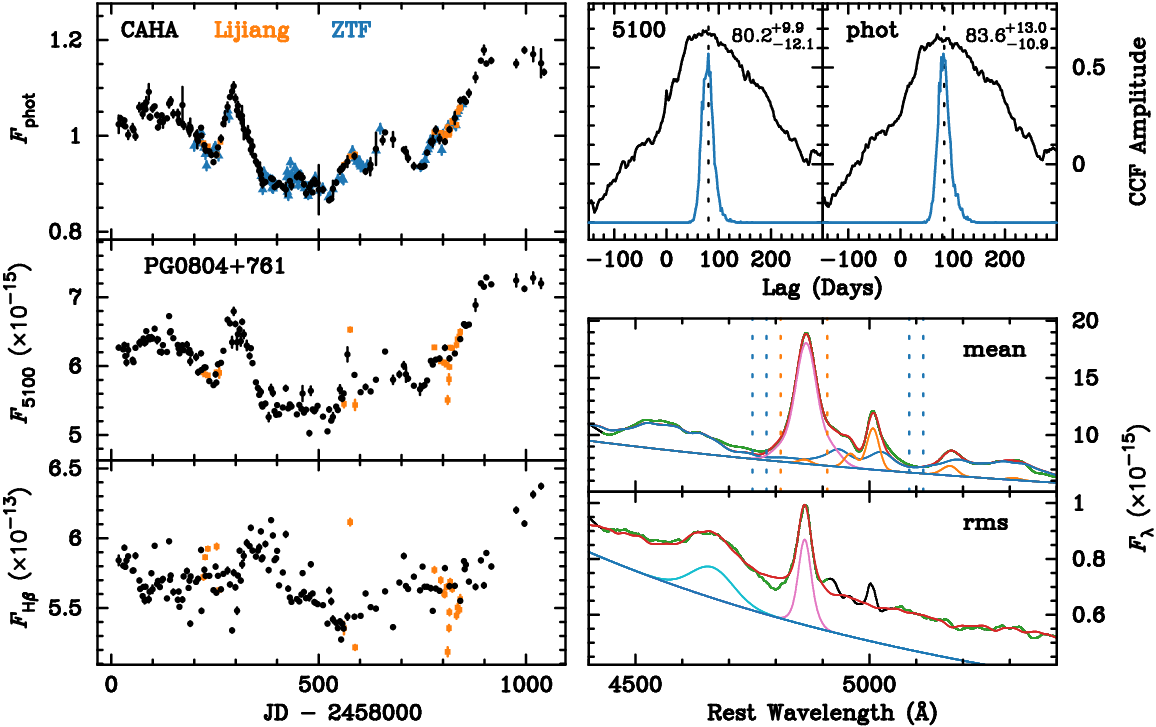}
  \caption{
  PG 0804+761 light curves, CCFs, mean and rms spectra. Same notations as in
  Figure \ref{fig-lc0003}.
  }
  \label{fig-lc0804}
\end{figure*}

\begin{figure*}
  \centering
  \includegraphics[width=0.9\textwidth]{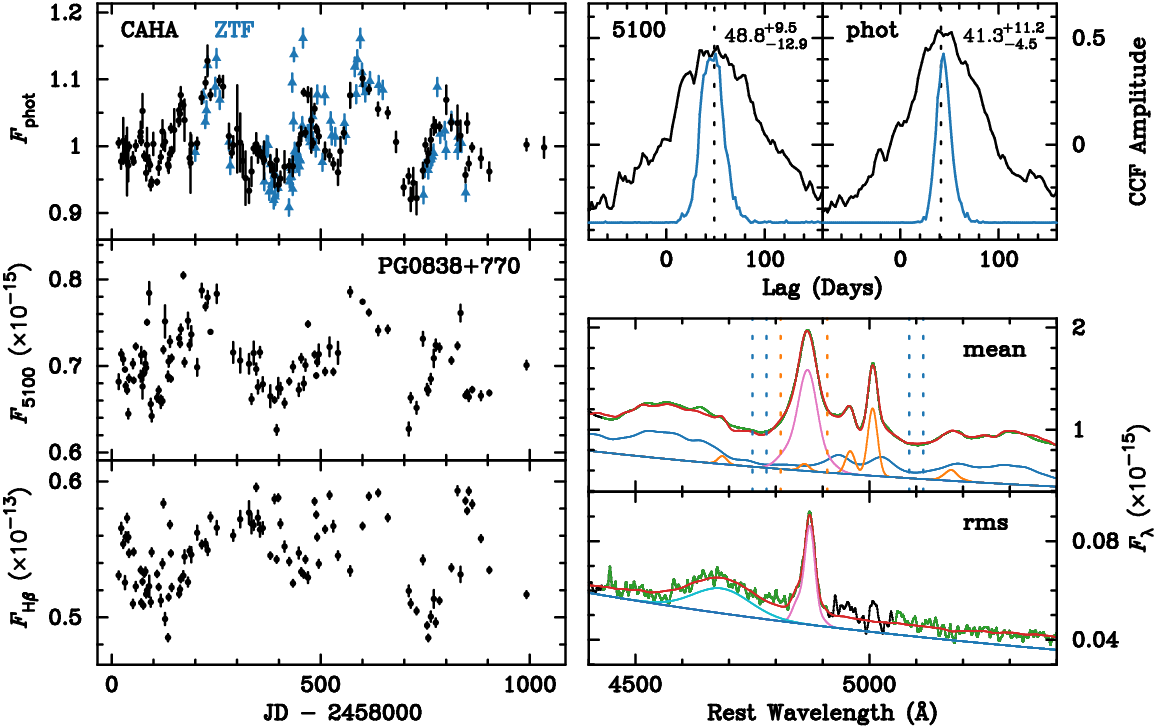}
  \caption{
  PG 0838+770 light curves, CCFs, mean and rms spectra. Same notations as in
  Figure \ref{fig-lc0003}.
  }
  \label{fig-lc0838}
\end{figure*}

\begin{figure*}
  \centering
  \includegraphics[width=0.9\textwidth]{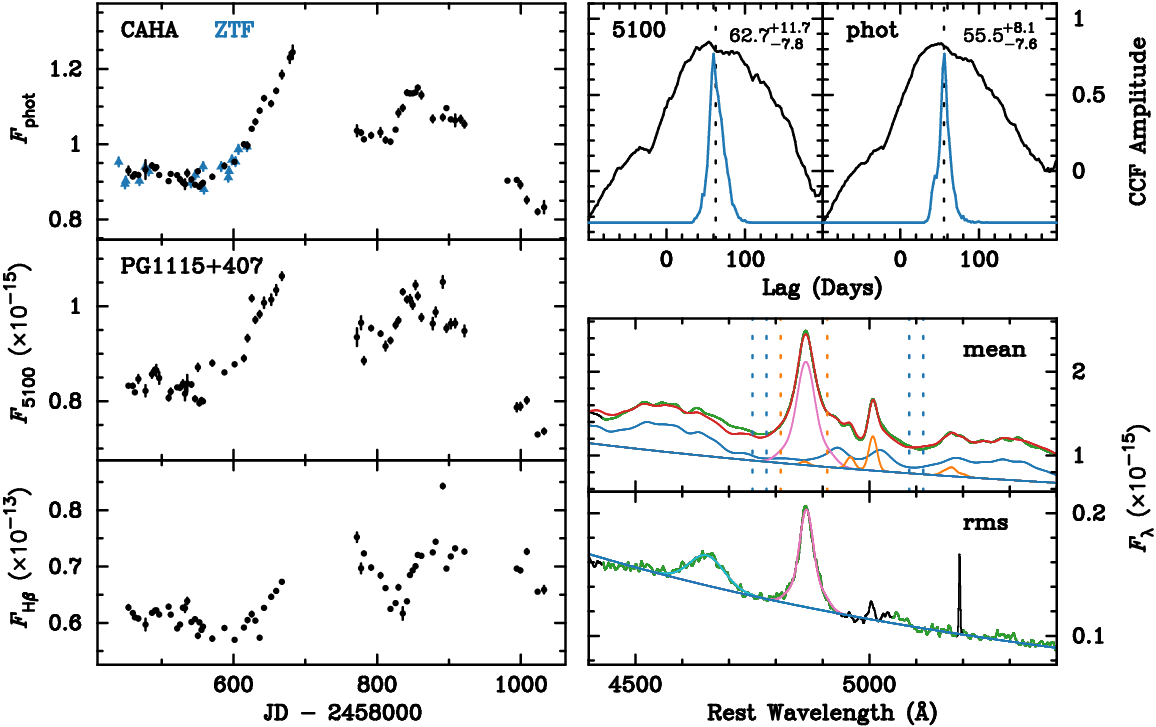}
  \caption{
  PG 1115+407 light curves, CCFs, mean and rms spectra. Same notations as in
  Figure \ref{fig-lc0003}.
  }
  \label{fig-lc1115}
\end{figure*}

\begin{figure*}
  \centering
  \includegraphics[width=0.9\textwidth]{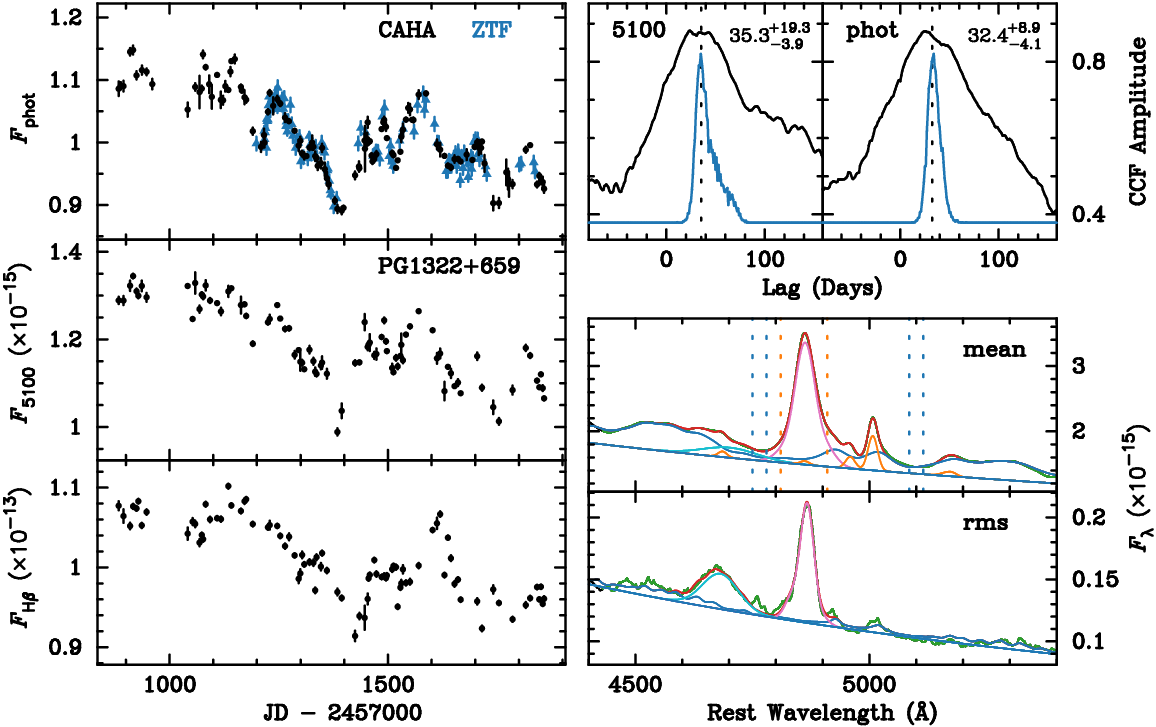}
  \caption{
  PG 1322+659 light curves, CCFs, mean and rms spectra. Same notations as in
  Figure \ref{fig-lc0003}.
  }
  \label{fig-lc1322}
\end{figure*}

\begin{figure*}
  \centering
  \includegraphics[width=0.9\textwidth]{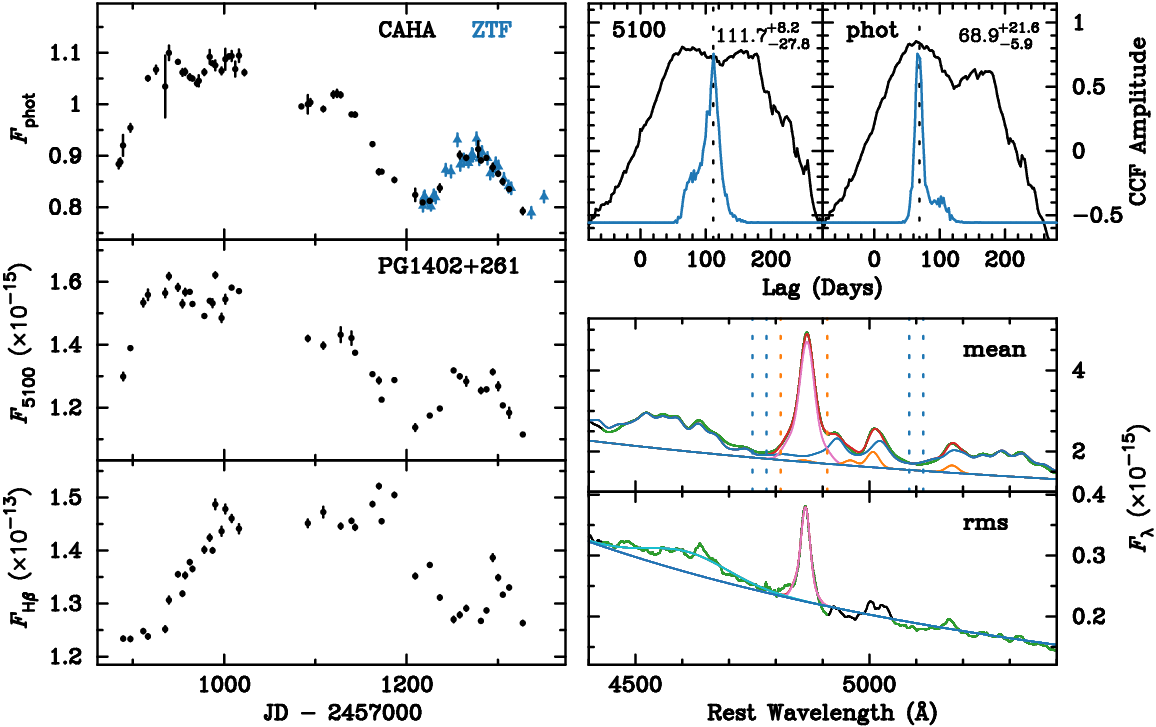}
  \caption{
  PG 1402+261 light curves, CCFs, mean and rms spectra. Same notations as in
  Figure \ref{fig-lc0003}.
  }
  \label{fig-lc1402}
\end{figure*}

\begin{figure*}
  \centering
  \includegraphics[width=0.9\textwidth]{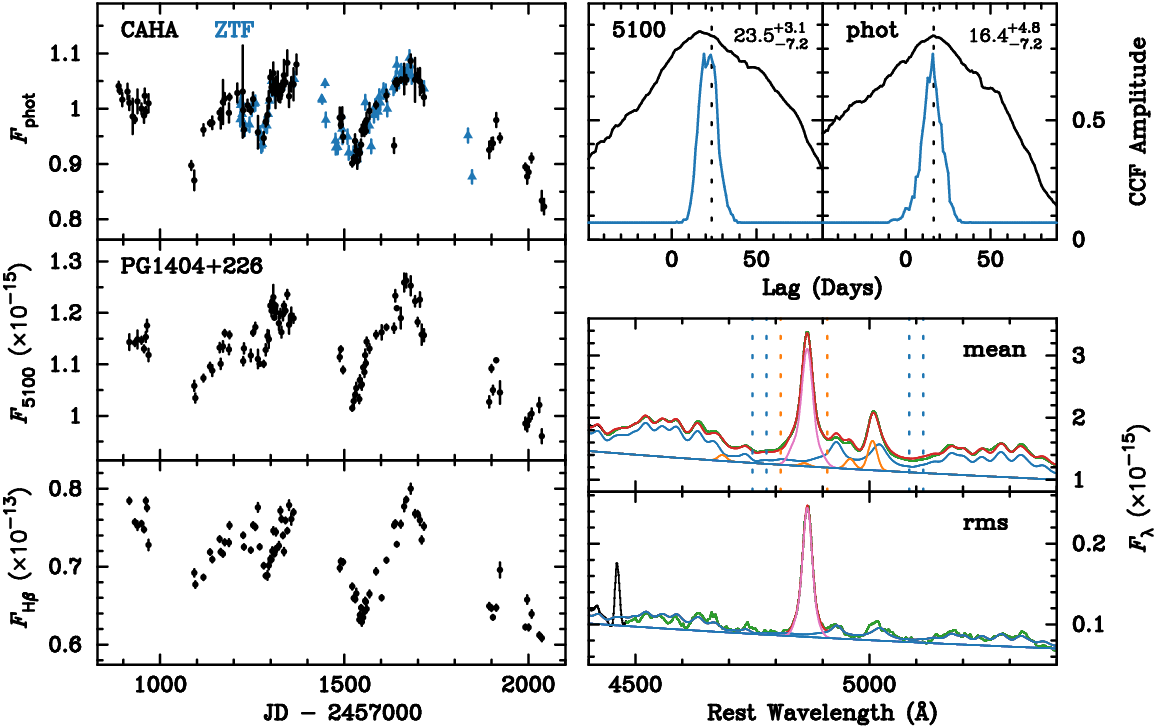}
  \caption{
  PG 1404+226 light curves, CCFs, mean and rms spectra. Same notations as in
  Figure \ref{fig-lc0003}.
  }
  \label{fig-lc1404}
\end{figure*}

\begin{figure*}
  \centering
  \includegraphics[width=0.9\textwidth]{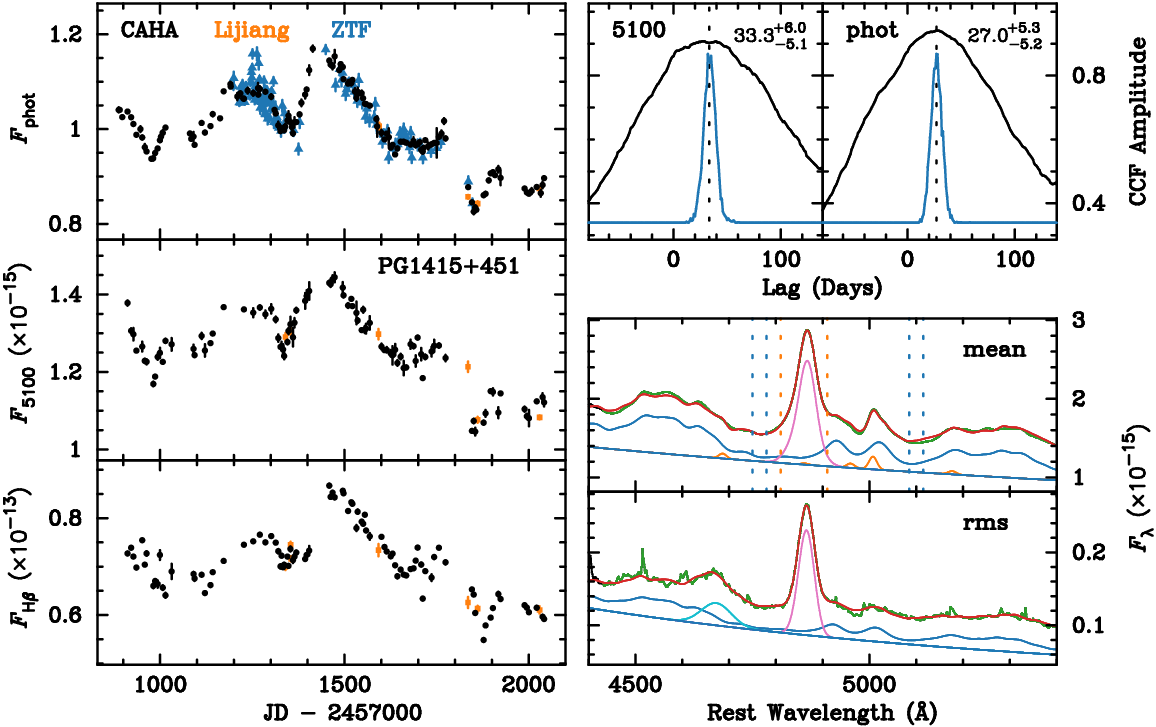}
  \caption{
  PG 1415+451 light curves, CCFs, mean and rms spectra. Same notations as in
  Figure \ref{fig-lc0003}.
  }
  \label{fig-lc1415}
\end{figure*}

\begin{figure*}
  \centering
  \includegraphics[width=0.9\textwidth]{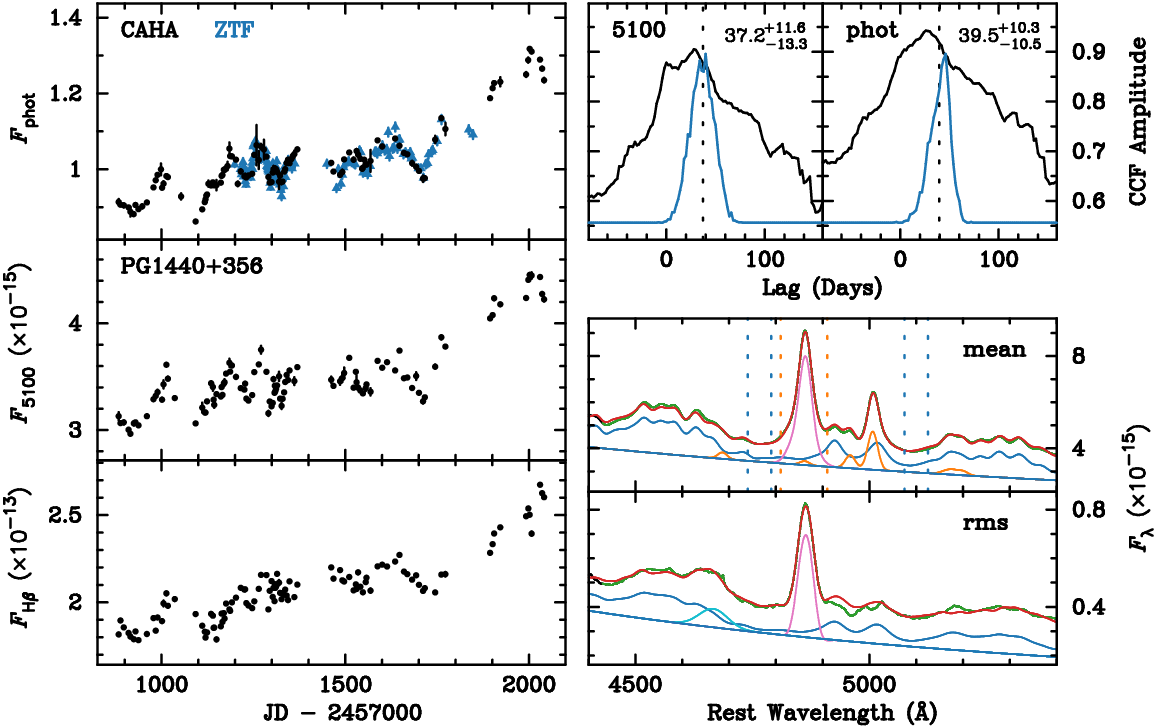}
  \caption{
  PG 1440+356 light curves, CCFs, mean and rms spectra. Same notations as in
  Figure \ref{fig-lc0003}.
  }
  \label{fig-lc1440}
\end{figure*}

\begin{figure*}
  \centering
  \includegraphics[width=0.9\textwidth]{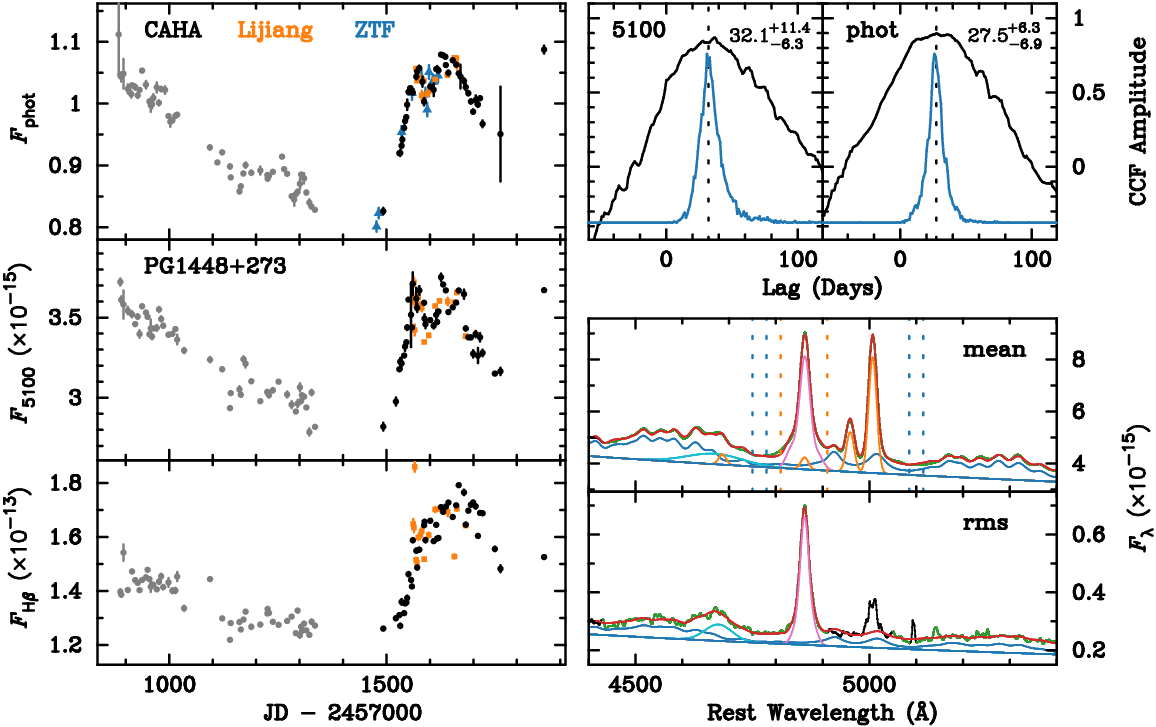}
  \caption{
  PG 1448+273 light curves, CCFs, mean and rms spectra. Same notations as in
  Figure \ref{fig-lc0003}. The gray data points are excluded in the
  time-series analysis, see the text in Section \ref{sec-note} for details.
  }
  \label{fig-lc1448}
\end{figure*}

\begin{figure*}
  \centering
  \includegraphics[width=0.9\textwidth]{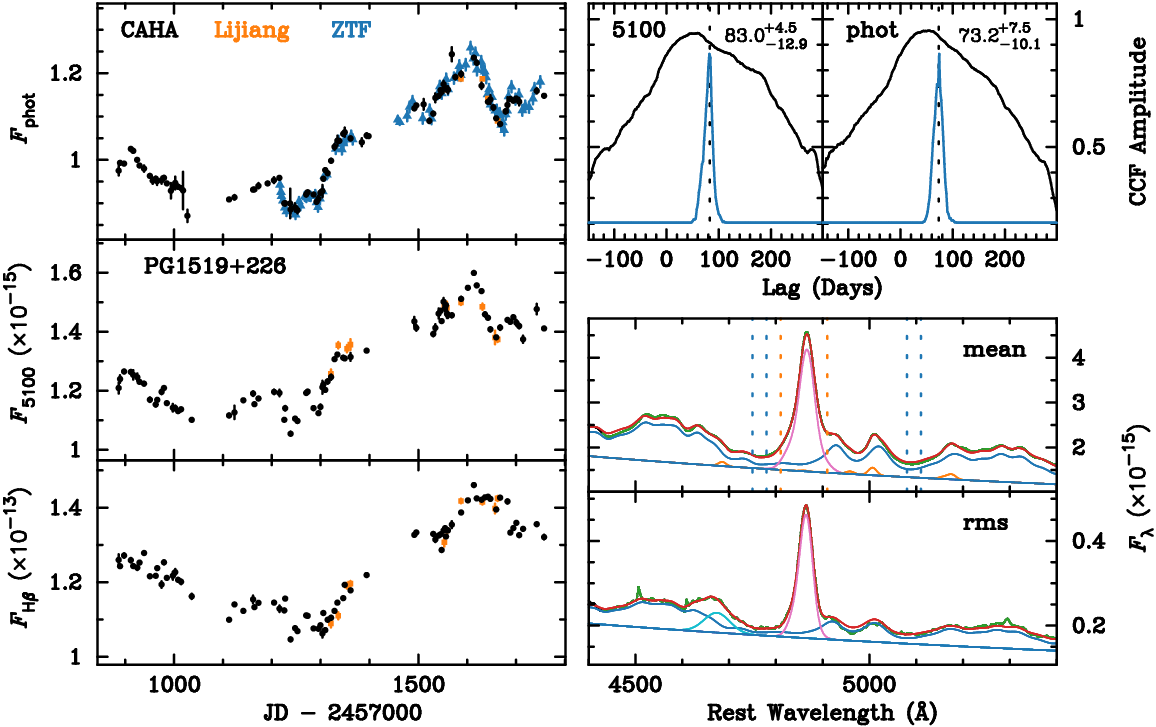}
  \caption{
  PG 1519+226 light curves, CCFs, mean and rms spectra. Same notations as in
  Figure \ref{fig-lc0003}.
  }
  \label{fig-lc1519}
\end{figure*}

\begin{figure*}
  \centering
  \includegraphics[width=0.9\textwidth]{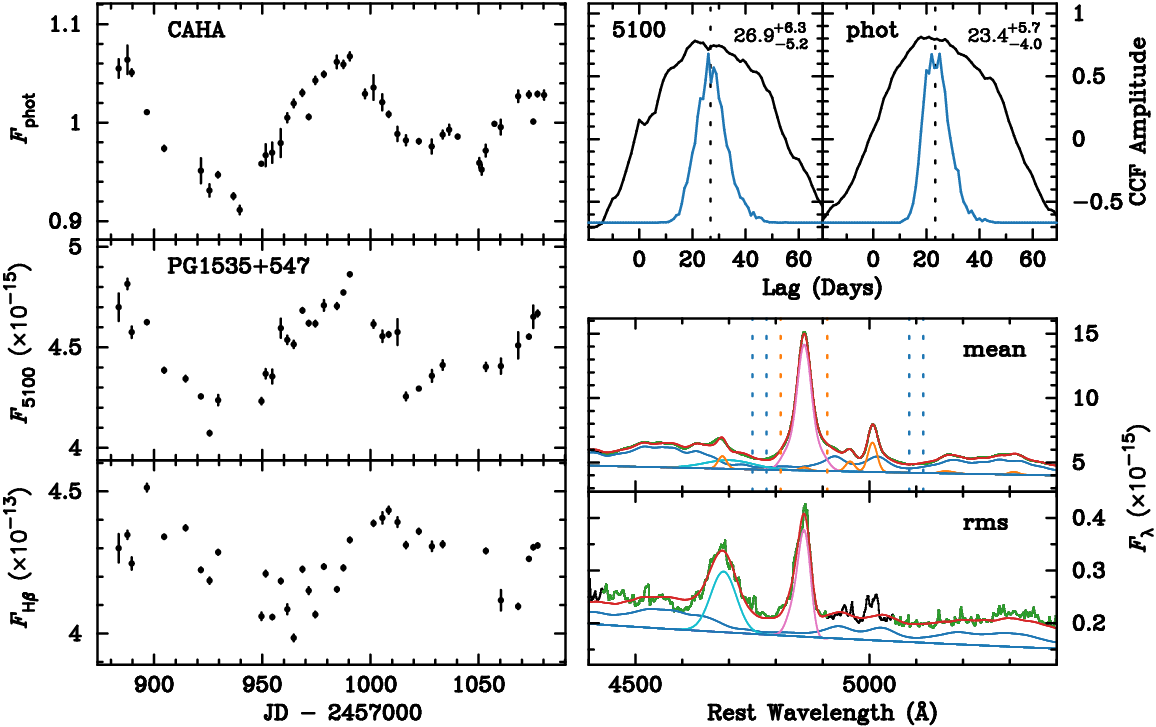}
  \caption{
  PG 1535+547 light curves, CCFs, mean and rms spectra. Same notations as in
  Figure \ref{fig-lc0003}.
  }
  \label{fig-lc1535}
\end{figure*}

\begin{figure*}
  \centering
  \includegraphics[width=0.9\textwidth]{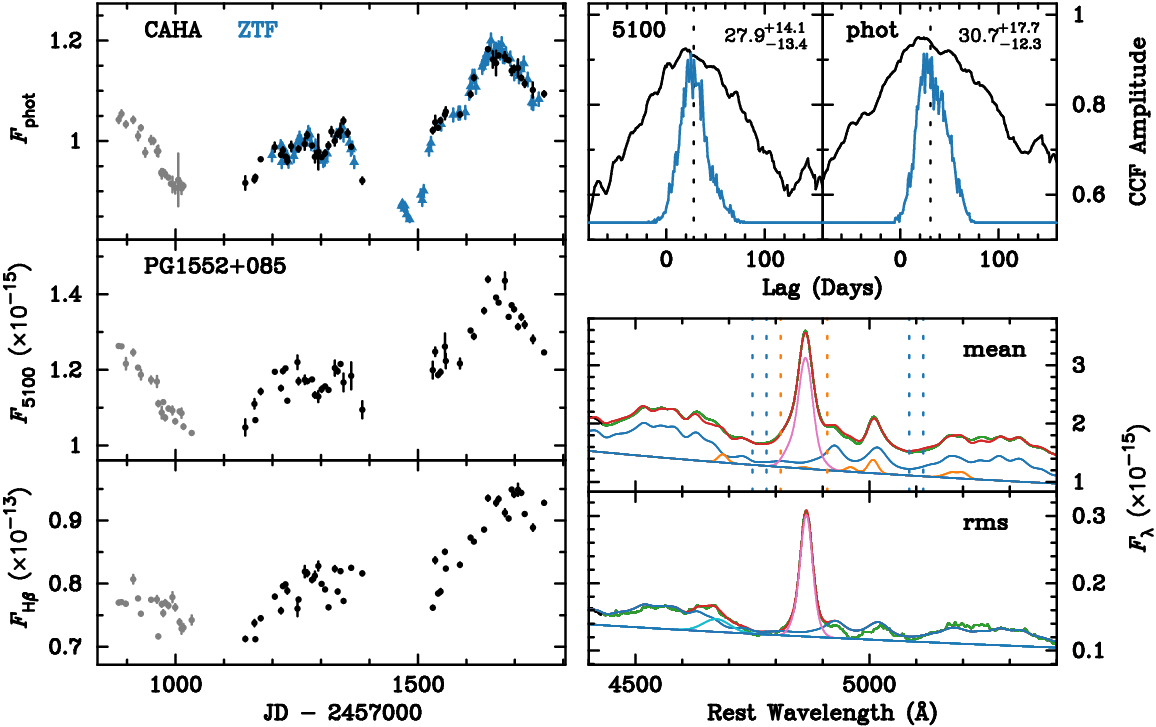}
  \caption{
  PG 1552+085 light curves, CCFs, mean and rms spectra. Same notations as in
  Figure \ref{fig-lc0003}. The gray data points are excluded in the
  time-series analysis, see the text in Section \ref{sec-note} for details.
  }
  \label{fig-lc1552}
\end{figure*}

\begin{figure*}
  \centering
  \includegraphics[width=0.9\textwidth]{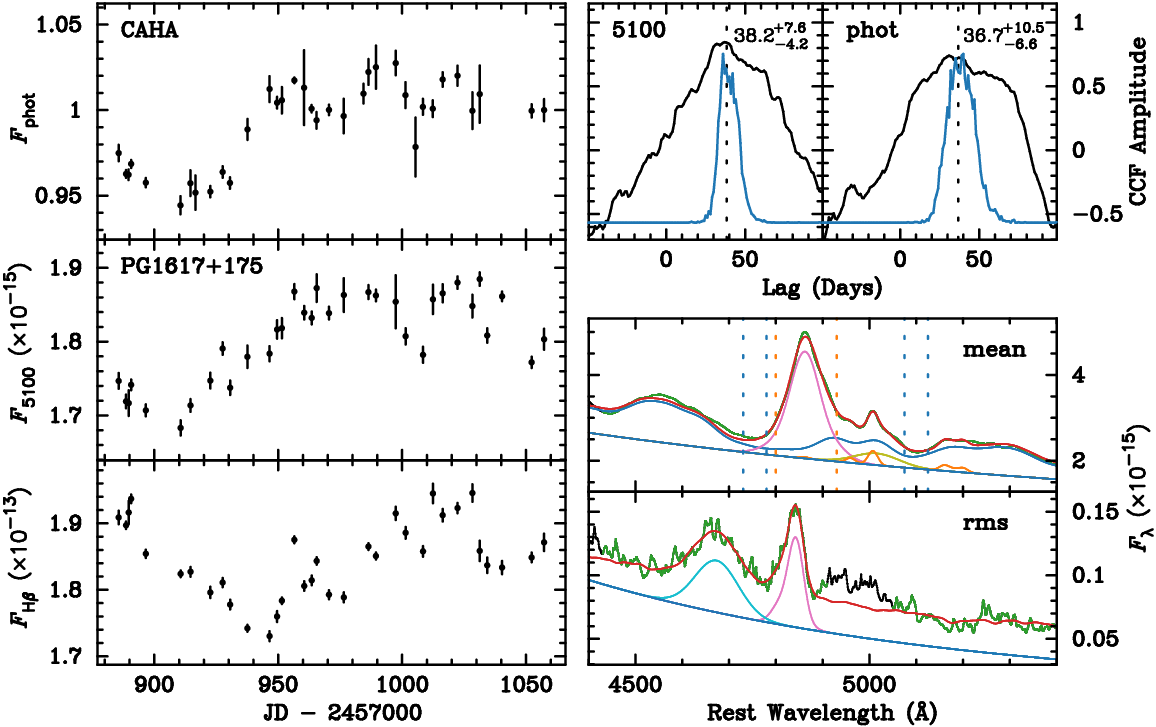}
  \caption{
  PG 1617+175 light curves, CCFs, mean and rms spectra. Same notations as in
  Figure \ref{fig-lc0003}.
  }
  \label{fig-lc1617}
\end{figure*}

\begin{figure*}
  \centering
  \includegraphics[width=0.9\textwidth]{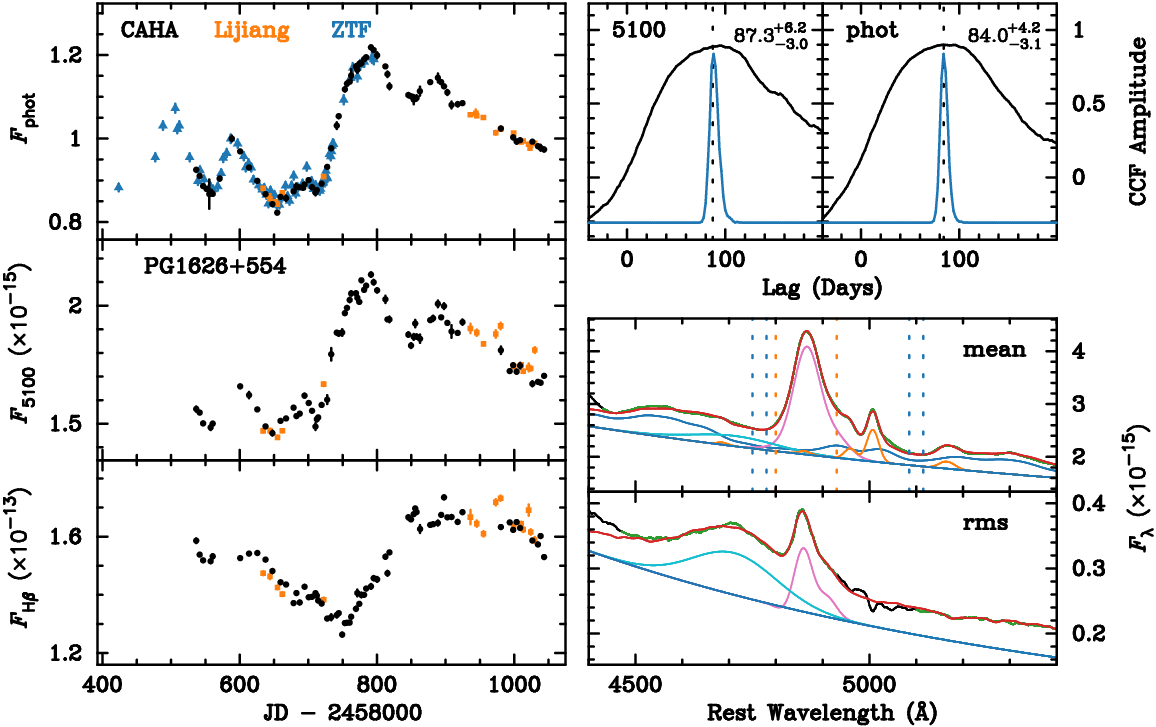}
  \caption{
  PG 1626+554 light curves, CCFs, mean and rms spectra. Same notations as in
  Figure \ref{fig-lc0003}.
  }
  \label{fig-lc1626}
\end{figure*}

The two bottom-right panels in Figures \ref{fig-lc0003}--\ref{fig-lc1626} show
the mean and rms spectra generated from the fully calibrated individual-epoch
spectra for each object in our sample, after applying a Galactic extinction
correction and converting to the rest frame using the values of $A_V$ and $z$
listed in Table \ref{tab-obs}, with the spectral decompositions (Section
\ref{sec-mass} below). For most objects, the \oiii\ lines are difficult to see
in the rms spectra, confirming the consistency of the flux calibration. There
might still be residual \oiii\ lines in the rms spectrum, as in the case of PG
0003+119 here (Figure \ref{fig-lc0003}), if the narrow-line region is strong
and extended. This could lead to differing slit losses for point (comparison
star, AGN continuum, and broad emission lines) and extended sources (narrow
emission lines and the host starlight), and may manifest itself in apparent
\oiii\ flux variations, an effect also noted in \citet{hu15} and \citet{lu19}.

\subsection{Photometry}
\label{sec-phot}

Broad-band photometric images were also taken by CAFOS in direct imaging mode
with a Johnson $V$ filter before the spectral exposures. The data were reduced
through standard IRAF procedures, and the fluxes were measured directly
within circular apertures. The size of the extraction aperture was adjusted
to include the entire host galaxy for each object. Then the differential
instrumental magnitudes were calculated for both the quasar and the comparison 
star with respect to other stars within the field of view.

All the comparison stars used in our observations were confirmed to be
non-varying during our campaign. The $V$-band light curves of the quasars are
shown as black data points in the top-left panels of Figures
\ref{fig-lc0003}--\ref{fig-lc1626}. The fluxes are plotted in arbitrary linear
units converted from the differential magnitudes, chosen to match the
photometric light curves obtained from other telescopes and bands (Section
\ref{sec-lcother} below).

\section{Light Curve Measurements}
\label{sec-lc}

Integration over a range of wavelengths above a continuum is the most widely
used method for defining the line fluxes used for BLR light curves in
reverberation mapping studies (e.g.,
\citealt{kaspi00,peterson04,bentz09,grier12,fausnaugh17,du18a,du18b}). It is
simple and robust for strong, single emission lines such as \ha\ and \hb, if
the contamination from the host galaxy \citep[e.g.,][]{hu16} and \heii\
\citep[e.g.,][]{hu20a} are not severe. The results obtained by the integration
method are consistent with those determined through spectral fitting under
normal situations (see Figure 6 of \citealt{hu15} for a comparison), or even a
little better in the case that the spectral quality is not good enough for
reliable decomposition of the host contribution \citep[e.g.,][]{hu20b}.
We performed input-output simulations to compare the performance and
suitability of two methods that can be utilized to determine the light curve:
integration and spectral fitting. We found that the integration method
performed more reliably for spectra of the typical quality and host galaxy
fractions encountered in this study, especially for the 5100 \AA\ continuum
light curve. See the Appendix for details. Thus, in this work, we use the
simple integration to measure the light curves of the 5100 \AA\ continuum and
the \hb\ broad emission line. Spectral fitting will be performed for
measuring those emission lines blended together, e.g., \heii\ and \feii, in a
forthcoming paper.

For each spectrum, a local linear continuum was defined by two continuum
windows near \hb. The red-ward one is located around the rest-frame 5100 \AA,
where the contamination of \feii\ is known to be weak. The mean flux in this
window is also used as the optical continuum flux (\fc). The blue-ward
continuum window is set at the local minimum between \hb\ and the bump of
blended \heii\ and \feii\ emission. The fluxes above this continuum were
integrated in a window around \hb\ $\lambda$4861 to measure the \hb\ flux. The
blue-ward boundary of this line window is selected to contain most of the
varying flux according to the rms spectrum, and the red-ward boundary is often
limited to avoid the contamination of \feii\ $\lambda$4924 emission. For each
object, the ranges of the line and continuum windows are marked by the orange
and blue dotted vertical lines, respectively, in the mean spectrum panel in
Figures \ref{fig-lc0003}--\ref{fig-lc1626}.

\begin{deluxetable}{lccr@{~$\pm$~}l}
  \tablewidth{0pt}
  \tablecaption{Light Curves
  \label{tab-lc}}
  \tablehead{
  \colhead{Object} & \colhead{Measure} & \colhead{JD} &
  \multicolumn{2}{c}{Flux} 
  \\
  \colhead{(1)} & \colhead{(2)} & \colhead{(3)} &
  \multicolumn{2}{c}{(4)}
  } 
  \startdata
0003+199 &  $V$ & 8693.612 & 2.182 & 0.005 \\
0003+199 &  $V$ & 8699.617 & 2.174 & 0.003 \\
\multicolumn{5}{c}{$\vdots$} \\
0003+199 & 5100 & 8693.623 & 4.301 & 0.008 \\
0003+199 & 5100 & 8699.629 & 4.259 & 0.015 \\
\multicolumn{5}{c}{$\vdots$} \\
0003+199 &  \hb & 8693.623 & 4.761 & 0.016 \\
0003+199 &  \hb & 8699.629 & 4.883 & 0.014 \\
\multicolumn{5}{c}{$\vdots$} \\
0804+761 &  $V$ & 8017.703 & 2.411 & 0.019 \\
0804+761 &  $V$ & 8022.677 & 2.398 & 0.006 \\
\multicolumn{5}{c}{$\vdots$} \\
0804+761 & 5100 & 8017.714 & 6.269 & 0.027 \\
0804+761 & 5100 & 8028.653 & 6.243 & 0.016 \\
\multicolumn{5}{c}{$\vdots$} \\
0804+761 &  \hb & 8017.714 & 5.841 & 0.034 \\
0804+761 &  \hb & 8028.653 & 5.804 & 0.012 \\
\multicolumn{5}{c}{$\vdots$}
  \enddata
  \tablecomments{
  This table is available in its entirety in a machine-readable form online.
  Example lines with the last few digits of JD are shown here. The fluxes
  are in units of instrumental magnitudes, $10^{-15}$ \ergscma, and $10^{-13}$
  \ergscm\ for the $V$ band, 5100 \AA, and \hb, respectively.
  }
\end{deluxetable}

The left panels in each of Figures \ref{fig-lc0003}--\ref{fig-lc1626} show
the light curves of \fc\ (middle panel) and \fhb\ (bottom panel), along with
the $V$ band light curve from CAHA (top panel; described in Section
\ref{sec-phot} above), as black data points, for each object. The data of
these three light curves are presented in its entirety online as a
machine-readable table. Table \ref{tab-lc} lists several example lines to
illustrate the structure of the online table. Note that \fc\ and \fhb\ are
measured in the observed frame, and without any Galactic-extinction
correction.

\subsection{Light Curves from Other Telescopes}
\label{sec-lcother}

During our campaign, some objects were observed for a few epochs by the
Lijiang 2.4m telescope at the Yunnan Observatory of the Chinese Academy of
Sciences. Both photometric and spectroscopic observations were carried out as
by \citet{du14} and \citet{hu20b}. The light curves were measured and
intercalibrated%
\footnote{
By applying a multiplicative and an additive factors as Equations (1) and (2)
in \citet{peterson02} to correct the systematic differences, e.g., aperture
effects, between different data sets.
}
to match those from CAHA, and plotted as orange squares in
Figures \ref{fig-lc0003}--\ref{fig-lc1626}. We also retrieved the light curves
from the Zwicky Transient Facility%
\footnote{\url{https://www.ztf.caltech.edu/}}
(ZTF) survey \citep{masci19} for most of our objects, shown as blue triangles
in Figures \ref{fig-lc0003}--\ref{fig-lc1626} after intercalibration.

The light curves from Lijiang and ZTF are consistent with the CAHA light
curves. However, the Lijiang observations are rather few, while the sampling
of ZTF during our campaign are not uniform. Thus, in the interest of
homogeneity, only the CAHA data set is used in the following time-series
measurements.

\section{Time-Series Measurements}
\label{sec-ccf}

\begin{deluxetable*}{lr@{~$\pm$~}lr@{~$\pm$~}lcr@{~$\pm$~}lcr@{~$\pm$~}lcr@{~$\pm$~}l}
  \tablewidth{0pt}
  \tablecaption{Light curve statistics
  \label{tab-var}}
  \tablehead{
  \colhead{Object} & \multicolumn{2}{c}{$V$} &
  \multicolumn{5}{c}{5100 \AA} & &
  \multicolumn{5}{c}{\hb}
  \\ \cline{4-8} \cline{10-14}
  \colhead{} & \multicolumn{2}{c}{\fvar} &
  \multicolumn{2}{c}{Flux} & \colhead{$\sigma_{\rm sys}$} &
  \multicolumn{2}{c}{\fvar} & &
  \multicolumn{2}{c}{Flux} & \colhead{$\sigma_{\rm sys}$} &
  \multicolumn{2}{c}{\fvar}
  \\ \cline{4-6} \cline{10-12}
  \colhead{} & \multicolumn{2}{c}{(\%)} &
  \multicolumn{3}{c}{($10^{-15}$ \ergscma)} & \multicolumn{2}{c}{(\%)} & &
  \multicolumn{3}{c}{($10^{-13}$ \ergscm)} & \multicolumn{2}{c}{{(\%)}}
  \\
  \colhead{(1)} & \multicolumn{2}{c}{(2)} &
  \multicolumn{2}{c}{(3)} & \colhead{(4)} & \multicolumn{2}{c}{(5)} & &
  \multicolumn{2}{c}{(6)} & \colhead{(7)} & \multicolumn{2}{c}{(8)} 
  } 
  \startdata
PG 0003+199 &  5.1 & 0.6 & 4.068 & 0.305 & 0.036 &  7.4 & 0.8 & & 4.474 & 0.195 & 0.045 & 4.2 & 0.5 \\
PG 0804+761 &  7.4 & 0.4 & 6.040 & 0.508 & 0.085 &  8.3 & 0.5 & & 5.702 & 0.189 & 0.088 & 2.9 & 0.2 \\
PG 0838+770 &  3.7 & 0.3 & 0.703 & 0.037 & 0.018 &  4.5 & 0.4 & & 0.545 & 0.027 & 0.016 & 4.0 & 0.4 \\
PG 1115+407 & 10.2 & 0.9 & 0.904 & 0.088 & 0.020 &  9.4 & 0.9 & & 0.650 & 0.056 & 0.018 & 8.2 & 0.8 \\
PG 1322+659 &  6.0 & 0.4 & 1.191 & 0.083 & 0.014 &  6.8 & 0.5 & & 1.011 & 0.045 & 0.010 & 4.3 & 0.3 \\
PG 1402+261 &  9.8 & 1.0 & 1.397 & 0.152 & 0.026 & 10.7 & 1.2 & & 1.369 & 0.086 & 0.021 & 6.1 & 0.7 \\
PG 1404+226 &  5.3 & 0.4 & 1.133 & 0.070 & 0.012 &  6.0 & 0.5 & & 0.711 & 0.049 & 0.011 & 6.6 & 0.5 \\
PG 1415+451 &  7.6 & 0.5 & 1.266 & 0.095 & 0.010 &  7.4 & 0.6 & & 0.712 & 0.068 & 0.012 & 9.4 & 0.7 \\
PG 1440+356 &  9.0 & 0.6 & 3.482 & 0.319 & 0.052 &  9.0 & 0.6 & & 2.069 & 0.184 & 0.035 & 8.7 & 0.6 \\
PG 1448+273 &  5.0 & 0.6 & 3.437 & 0.204 & 0.047 &  5.7 & 0.7 & & 1.563 & 0.153 & 0.045 & 9.3 & 1.1 \\
PG 1519+226 & 10.0 & 0.8 & 1.297 & 0.144 & 0.009 & 11.0 & 0.9 & & 1.243 & 0.115 & 0.014 & 9.2 & 0.7 \\
PG 1535+547 &  3.9 & 0.4 & 4.512 & 0.186 & 0.071 &  3.8 & 0.5 & & 4.252 & 0.121 & 0.063 & 2.4 & 0.4 \\
PG 1552+085 &  7.3 & 0.7 & 1.226 & 0.096 & 0.013 &  7.7 & 0.8 & & 0.831 & 0.067 & 0.016 & 7.9 & 0.9 \\
PG 1617+175 &  2.3 & 0.3 & 1.805 & 0.059 & 0.017 &  3.0 & 0.4 & & 1.848 & 0.058 & 0.017 & 3.0 & 0.4 \\
PG 1626+554 & 11.9 & 1.0 & 1.788 & 0.207 & 0.020 & 11.5 & 1.0 & & 1.502 & 0.129 & 0.019 & 8.5 & 0.8
  \enddata
  \tablecomments{
  Fluxes are the means and standard deviations in the light curves. The
  estimated systematic error $\sigma_{\rm sys}$ for each light curve has been
  included when calculating the \fvar\ and the uncertainty of the time lag.
  }
\end{deluxetable*}

The uncertainties in the fluxes listed in Table \ref{tab-lc} and plotted in
Figures \ref{fig-lc0003}--\ref{fig-lc1626} are the statistical errors given by
the light-curve measurements according to the errors in the observed counts.
In most cases, additional systematic errors introduced by, e.g., the method of
flux calibration, slit losses, and host contamination, are needed to interpret
the scatter in the light curves. By the same method as in \citet{hu20b}, we
estimated a systematic error for each light curve according to the differences
in the fluxes of successive epochs, as listed in columns (4) and (7) of Table
\ref{tab-var}. These systematic errors were then added in quadrature with the
statistical errors for the time-series measurements below.

\subsection{Variability Amplitudes}

We calculated the quantity \fvar\ \citep{rodriguez97} and its uncertainty
\citep{edelson02} to represent the intrinsic variability amplitude.
It is defined as
\begin{equation}
  F_{\rm var} = \frac{1}{\langle F \rangle} \sqrt{\sigma^2 - \langle
  \sigma_{\rm err}^2 \rangle}~,
\end{equation}
where $\sigma^2$ is the flux variance, $\langle \sigma_{\rm err}^2 \rangle$ is
the mean square error, and $\langle F \rangle$ is the mean
flux. And the uncertainty of \fvar\ is
\begin{equation}
  \sigma_{F_{\rm var}} = \frac{1}{F_{\rm var}} \sqrt{\frac{1}{2N}}
  \frac{\sigma^2}{\langle F \rangle^2}~,
\end{equation}
where $N$ is the number of epochs. The results are listed in columns (2),
(5), and (8) of Table \ref{tab-var} for the $V$-band, \fc, and \fhb\ light
curves, respectively. The \fvar\ of the photometric and spectral continuum in
this sample are consistent with each other, ranging from $\sim$3\% to
$\sim$11\%. The \fvar\ for \hb\ are on average smaller than those for the
continuum. The mean fluxes of the 5100 \AA\ continuum and \hb\ are also listed
as columns (3) and (6), respectively. The uncertainties are the standard
deviations of these means.

\subsection{Reverberation lags}
\label{sec-lag}

\begin{deluxetable}{lcr@{}lccr@{}l}
  \tablewidth{0pt}
  \tablecaption{Cross-Correlation Results
  \label{tab-ccf}}
  \tablehead{
  \colhead{Object} & \multicolumn{3}{c}{\hb\ vs. 5100 \AA} & &
  \multicolumn{3}{c}{\hb\ vs. $V$}
  \\ \cline{2-4} \cline{6-8}
  \colhead{} & \colhead{\rmax} & \multicolumn{2}{c}{Lag} & &
  \colhead{\rmax} & \multicolumn{2}{c}{Lag}
  \\
  \colhead{} & \colhead{} & \multicolumn{2}{c}{(days)} & &
  \colhead{} & \multicolumn{2}{c}{(days)}
  \\
  \colhead{(1)} & \colhead{(2)} & \multicolumn{2}{c}{(3)} & &
  \colhead{(4)} & \multicolumn{2}{c}{(5)}
  } 
  \startdata
PG 0003+199 & 0.69 & 17.0 & $_{ -3.2}^{+ 2.5}$ & & 0.65 & 16.4 & $_{ -3.8}^{+ 2.8}$  \\
PG 0804+761 & 0.69 & 72.9 & $_{-11.0}^{+ 9.0}$ & & 0.67 & 76.0 & $_{ -9.9}^{+11.8}$  \\
PG 0838+770 & 0.46 & 43.1 & $_{-11.4}^{+ 8.4}$ & & 0.53 & 36.5 & $_{ -3.9}^{+ 9.9}$  \\
PG 1115+407 & 0.85 & 54.4 & $_{ -6.7}^{+10.2}$ & & 0.84 & 48.1 & $_{ -6.6}^{+ 7.0}$  \\
PG 1322+659 & 0.88 & 30.3 & $_{ -3.3}^{+16.5}$ & & 0.88 & 27.7 & $_{ -3.6}^{+ 7.6}$  \\
PG 1402+261 & 0.81 & 95.9 & $_{-23.9}^{+ 7.1}$ & & 0.85 & 59.1 & $_{ -5.0}^{+18.5}$  \\
PG 1404+226 & 0.87 & 21.4 & $_{ -6.5}^{+ 2.8}$ & & 0.85 & 15.0 & $_{ -6.6}^{+ 4.3}$  \\
PG 1415+451 & 0.91 & 30.0 & $_{ -4.6}^{+ 5.4}$ & & 0.94 & 24.3 & $_{ -4.7}^{+ 4.8}$  \\
PG 1440+356 & 0.91 & 34.5 & $_{-12.4}^{+10.8}$ & & 0.94 & 36.7 & $_{ -9.8}^{+ 9.6}$  \\
PG 1448+273 & 0.87 & 30.1 & $_{ -5.9}^{+10.7}$ & & 0.90 & 25.8 & $_{ -6.5}^{+ 5.9}$  \\
PG 1519+226 & 0.95 & 73.1 & $_{-11.4}^{+ 4.0}$ & & 0.95 & 64.5 & $_{ -8.9}^{+ 6.6}$  \\
PG 1535+547 & 0.78 & 25.9 & $_{ -5.0}^{+ 6.1}$ & & 0.81 & 22.5 & $_{ -3.8}^{+ 5.5}$  \\
PG 1552+085 & 0.92 & 25.0 & $_{-12.0}^{+12.6}$ & & 0.95 & 27.5 & $_{-11.0}^{+15.9}$  \\
PG 1617+175 & 0.85 & 34.3 & $_{ -3.8}^{+ 6.8}$ & & 0.74 & 33.0 & $_{ -5.9}^{+ 9.4}$  \\
PG 1626+554 & 0.89 & 77.1 & $_{ -2.6}^{+ 5.5}$ & & 0.90 & 74.3 & $_{ -2.7}^{+ 3.7}$ 
  \enddata
  \tablecomments{
  Lags are in the rest frame.
  }
\end{deluxetable}

We calculated the reverberation time lags between the continuum and
emission-line variations using the standard interpolation cross-correlation
function (CCF) method \citep{gaskell86,gaskell87,white94}. The centroid of the
CCF above the 80\% level of the peak value (\rmax) was adopted following
\citet{koratkar91} and \citet{peterson04}. Monte Carlo simulations of random
subset selection (RSS) and flux randomization \citep{maoz89,peterson98b} were
performed to obtain the cross-correlation centroid distribution (CCCD), from
which the uncertainty of the time lag was estimated.

The two top-right panels in Figures \ref{fig-lc0003}--\ref{fig-lc1626} show
the CCFs (in black) and corresponding CCCDs (in blue) for \hb\ with respect to
the spectral 5100 \AA\ and $V$-band continuum, respectively. The calculated
rest-frame time lags of the \hb\ with respect to both the photometric
continuum (\hb\ versus $V$; \tph) and the spectral continuum (\hb\ versus
5100; \tsp) are listed in Table \ref{tab-ccf}, along with the corresponding
\rmax. In principle, the $V$-band flux could be contaminated by emission lines
to some degree, depending on the redshift of the object. Thus, \tsp\ is
adopted for the following analysis and calculations. Actually, the values of
\tph\ and \tsp\ are very close to each other in all objects, except PG
1402+261, whose CCFs show double-peaked profiles as noted below.

\subsubsection{Notes to individual objects}
\label{sec-note}

\textit{PG 0838+770}. The light curves of \fc\ and \fhb\ in Figure
\ref{fig-lc0838} appear to have more scatter than indicated by the error bars.
The estimated additional systematic errors (listed in Table \ref{tab-var}) are
$\sim$2.6\% and $\sim$3.0\% of the mean fluxes for \fc\ and \fhb,
respectively. Both are larger percentage-wise than for any other object in our
sample. The contribution of the host galaxy starlight, which is easily seen in
the spatial profile when extracting the spectrum, may lead to scatter in the
light curves for its apparent flux variations due to changing observing
conditions (\citealt{hu15}; see also the simulations in the Appendix). The
CCFs have \rmax\ of only $\sim$0.5 (Table \ref{tab-ccf}), and the RSS
simulations of its rms spectrum yield a large uncertainty in the velocity
width of the \hb\ emission line (see Section \ref{sec-mass} below).

\textit{PG 1402+261}. The two peaks in the CCF with respect to \fc\ are both
higher than 80\% of \rmax, and the centroid is between the peaks (see Figure
\ref{fig-lc1402}). However, the $\sim$160d peak is significantly lower than
the one at $\sim$80d in the CCF with respect to \fph, so the centroid is
consistent with the peak associated with the shorter period. Hence we find
that \tph\ $\approx$ 2/3\tsp, but these two lags are still consistent with one
another considering the large lower uncertainty in \tsp\ ($95.9_{-23.9}^{+
7.1}$ days) as well as the large upper uncertainty in \tph\ ($59.1_{
-5.0}^{+18.5}$ days). 

\textit{PG 1448+273}. The long-term trend of the \fhb\ light curve appears to
differ from both continuum light curves (Figure \ref{fig-lc1448}). The broad
\heii\ emission line is weak in both mean and rms spectra, indicating that the
trend is not caused by the contamination of \heii\ in the integration as in
the case of PG 0026+129 (\citealt{hu20a}, Figures 2 and 3). The measured time
lag can be biased by these long-term trends \citep{welsh99}, and detrending
(subtract a low-order polynomial to remove the long-term trend) were applied
in previous studies sometimes \citep[e.g.,][]{fausnaugh17,zhang19}. However,
such a detrending is somewhat arbitrary and artificial without the knowledge
of the physical interpretation to the long-term trend \citep{li20},
introducing additional uncertainty in the measured time lag. The light curves
before JD 2458400 (marked in gray) display little more than a steady decline,
making it impractical to use these to determine CCFs. These values were
therefore excluded from the CCF calculation. In this way, the influence of the
long-term trend is avoided for the remaining data in just one season.

\textit{PG 1552+085}. Similar to the above case of PG 1448+273, the light
curves before JD 2458100 (gray data points in Figure \ref{fig-lc1552}) are
excluded in the CCF calculations to alleviate the uncertainty introduced
by the different long-term trends in the continuum and the \hb\ light curves.

\section{Black Hole Masses}
\label{sec-mass}

Assuming that the motion of the BLR clouds are dominated by the gravitational
attraction of the central black hole, the latter's mass \mbh\ can be estimated
by
\begin{equation}
  M_{\rm BH} = f \frac{c \tau \Delta V^2}{G}~,
\end{equation}
where $c$ is the speed of light, $G$ is the gravitational constant, $\tau$ and
$\Delta V$ are the measured time lag and velocity width of the emission line
respectively, and $f$ is a dimensionless virial factor. For the $\tau$, we
adopt the time lag of \fhb\ with respect to \fc\ as mentioned in Section
\ref{sec-lag} above.

In practice, $\Delta V$ can be measured from either mean or rms spectra, and
characterized by either the FWHM or the line dispersion (\sline). The rms
spectrum is preferred here because it represents the varying part of the
emission line, which is the component whose responsivity-weighted-mean radius
is measured by the time lag. The rms spectrum also has the advantage of
removing the contamination of the unvarying narrow line. Following
\citet{peterson04}, we adopt the line dispersion in the rms spectrum,
\sline(rms), for the mass estimation.

\begin{deluxetable*}{lr@{~$\pm$~}lr@{~$\pm$~}lr@{~$\pm$~}lr@{~$\pm$~}lr@{}lr@{}lcr@{~$\pm$~}l}
  \tablewidth{0pt}
  \tablecaption{Line Widths, Virial Masses, and Luminosities
  \label{tab-mass}}
  \tablehead{
  \colhead{Object} & \multicolumn{2}{c}{FWHM(mean)} &
  \multicolumn{2}{c}{\sline(mean)} & \multicolumn{2}{c}{FWHM(rms)} &
  \multicolumn{2}{c}{\sline(rms)} & \multicolumn{2}{c}{Virial Product} &
  \multicolumn{2}{c}{\mbh} &
  \colhead{$\lambda L_{\rm \lambda,gal}$(5100 \AA)} &
  \multicolumn{2}{c}{$\lambda L_{\rm \lambda,AGN}$(5100 \AA)}
  \\
  \colhead{} & \multicolumn{2}{c}{(\kms)} &
  \multicolumn{2}{c}{(\kms)} & \multicolumn{2}{c}{(\kms)} &
  \multicolumn{2}{c}{(\kms)} & \multicolumn{2}{c}{($\times 10^7 M_\odot$)} &
  \multicolumn{2}{c}{($\times 10^7 M_\odot$)} &
  \colhead{($\times 10^{44}$ \ergs)} &
  \multicolumn{2}{c}{($\times 10^{44}$ \ergs)}
  \\
  \colhead{(1)} & \multicolumn{2}{c}{(2)} &
  \multicolumn{2}{c}{(3)} & \multicolumn{2}{c}{(4)} &
  \multicolumn{2}{c}{(5)} & \multicolumn{2}{c}{(6)} &
  \multicolumn{2}{c}{(7)} & \colhead{(8)} & \multicolumn{2}{c}{(9)} 
  } 
  \startdata
PG 0003+199 & 1748 & 120 & 1259 & 60 &  836 & 348 &  589 &  98 & 0.12 & $_{-0.04}^{+0.04}$ &  0.50 & $_{-0.19}^{+0.18}$ & 0.06 & 0.27 & 0.03 \\
PG 0804+761 & 3251 &  81 & 2007 & 24 & 1534 & 121 &  822 &  74 & 0.96 & $_{-0.23}^{+0.21}$ &  4.14 & $_{-0.98}^{+0.91}$ & 0.02 & 8.24 & 0.75 \\
PG 0838+770 & 2986 & 119 & 1724 & 35 & 1167 & 210 &  893 & 129 & 0.67 & $_{-0.26}^{+0.23}$ &  2.89 & $_{-1.13}^{+1.01}$ & 0.63 & 1.15 & 0.10 \\
PG 1115+407 & 2770 &  77 & 1709 & 14 & 2022 & 245 & 1303 & 143 & 1.80 & $_{-0.45}^{+0.52}$ &  7.76 & $_{-1.95}^{+2.23}$ & 0.75 & 2.43 & 0.33 \\
PG 1322+659 & 3032 &  66 & 1719 & 13 & 1846 &  54 & 1147 & 105 & 0.78 & $_{-0.17}^{+0.45}$ &  3.35 & $_{-0.71}^{+1.92}$ & 0.00 & 5.07 & 0.38 \\
PG 1402+261 & 2056 &  12 & 1368 &  9 &  772 & 339 &  650 & 119 & 0.79 & $_{-0.35}^{+0.30}$ &  3.41 & $_{-1.51}^{+1.28}$ & 0.00 & 5.49 & 0.66 \\
PG 1404+226 & 1303 &  14 &  985 & 16 &  839 & 114 &  612 &  50 & 0.16 & $_{-0.05}^{+0.03}$ &  0.68 & $_{-0.23}^{+0.14}$ & 0.15 & 1.28 & 0.09 \\
PG 1415+451 & 2473 &  23 & 1258 &  3 & 1989 &  65 &  833 &  42 & 0.41 & $_{-0.07}^{+0.08}$ &  1.75 & $_{-0.32}^{+0.36}$ & 0.47 & 1.70 & 0.17 \\
PG 1440+356 & 1868 &  48 & 1043 & 12 & 2032 &  51 &  715 &  37 & 0.34 & $_{-0.13}^{+0.11}$ &  1.49 & $_{-0.55}^{+0.49}$ & 0.44 & 2.07 & 0.25 \\
PG 1448+273 & 1323 & 139 & 1043 & 70 &  893 & 104 &  631 &  39 & 0.23 & $_{-0.05}^{+0.09}$ &  1.01 & $_{-0.23}^{+0.38}$ & 0.12 & 1.72 & 0.11 \\
PG 1519+226 & 2212 &  12 & 1315 &  3 & 1714 &  98 &  889 &  38 & 1.13 & $_{-0.20}^{+0.11}$ &  4.87 & $_{-0.86}^{+0.49}$ & 0.37 & 3.10 & 0.42 \\
PG 1535+547 & 2004 &  42 & 1311 & 16 & 1737 & 337 &  845 & 207 & 0.36 & $_{-0.19}^{+0.20}$ &  1.55 & $_{-0.82}^{+0.84}$ & 0.05 & 0.70 & 0.03 \\
PG 1552+085 & 1963 &  24 & 1193 &  5 & 1408 &  98 &  788 &  53 & 0.30 & $_{-0.15}^{+0.16}$ &  1.30 & $_{-0.65}^{+0.68}$ & 0.54 & 1.95 & 0.21 \\
PG 1617+175 & 4807 &  46 & 2511 &  3 & 2392 & 187 & 1288 & 374 & 1.11 & $_{-0.66}^{+0.68}$ &  4.79 & $_{-2.83}^{+2.94}$ & 0.12 & 2.96 & 0.12 \\
PG 1626+554 & 4318 &  95 & 2350 & 11 & 2784 & 180 & 1719 & 119 & 4.45 & $_{-0.63}^{+0.69}$ & 19.17 & $_{-2.73}^{+2.98}$ & 0.26 & 4.04 & 0.52
  \enddata
  \tablecomments{
  All widths have been corrected for the instrumental broadening. The virial
  products are calculated adopting the line dispersions in the rms spectra.
  The masses of the central black holes are estimated using the virial factor
  $f$=4.31 given by \citet{grier13}. The optical luminosities of the host
  galaxies and the AGNs are calculated from the fluxes of corresponding
  components in the mean spectra decomposed by spectral fitting, while the
  errors of the AGN luminosities are derived from the standard deviations in
  the light curves of the 5100 \AA\ continuum fluxes.
  }
\end{deluxetable*}

For completeness, FWHM and \sline\ were measured in both the mean and the rms
spectra, by using spectral fitting to decompose the broad \hb\ component (see
the bottom panels in the right column of Figures
\ref{fig-lc0003}--\ref{fig-lc1626}). The following components were included in
the fit to the mean spectra as in \citet{hu15}: (1) a single power law as
the AGN continuum, (2) the host galaxy modeled by single simple stellar
population template from \citet{bruzual03}, (3) \feii\ emission modeled by
the template from \citet{boroson92}, (4) a Gauss-Hermite function for the
broad \hb\ component, (5) a set of Gaussians with the same widths and shifts
for the narrow emission lines, and (6) a Gaussian for the broad \heii\
$\lambda$4686 emission line if necessary. Because of the relatively narrow
\hb\ broad component and weak narrow lines in our sample, and the wide
instrumental broadening of our spectra, it is not practical to decompose the
narrow \hb\ component through free fitting. Thus, following \citet{hu15}, we
fit the mean spectra assuming a flux ratio of 10\% for the narrow \hb\
component with respect to \oiii\ $\lambda$5007, and the resulting FWHM and
\sline\ of the broad \hb\ were adopted for the following analysis. Another two
fits with the flux ratio set to 0 and 20\% were also performed to estimate the
uncertainties in the widths.

Note that none of the rms spectra in our sample show complex multi-peaked
profiles of the \hb\ emission line (see those in Figures
\ref{fig-lc0003}--\ref{fig-lc1626}), thus the broad \hb\ emission line was
simply also modeled by a Gauss-Hermite function in the fit to the rms
spectrum. Same spectral components were included as in the fit to the mean
spectrum above, except the narrow emission lines. For a few objects with
residual \oiii\ lines (see Section \ref{sec-spec} for an explanation), the
wavelength region around were masked in the fitting. Note that the \feii\
emission are generally weaker in the rms spectra than in the mean spectra,
while the broad \heii\ lines are commonly much stronger. The RSS simulations
for the time lag uncertainty estimation are also used here to generate
realizations of rms spectra. The distributions of the FWHM and \sline\
measured from these realizations determine the uncertainties. The results,
after instrumental broadening correction, are listed in columns (2)--(5) of
Table \ref{tab-mass}. 

\begin{figure*}
  \centering
  \includegraphics[width=0.9\textwidth]{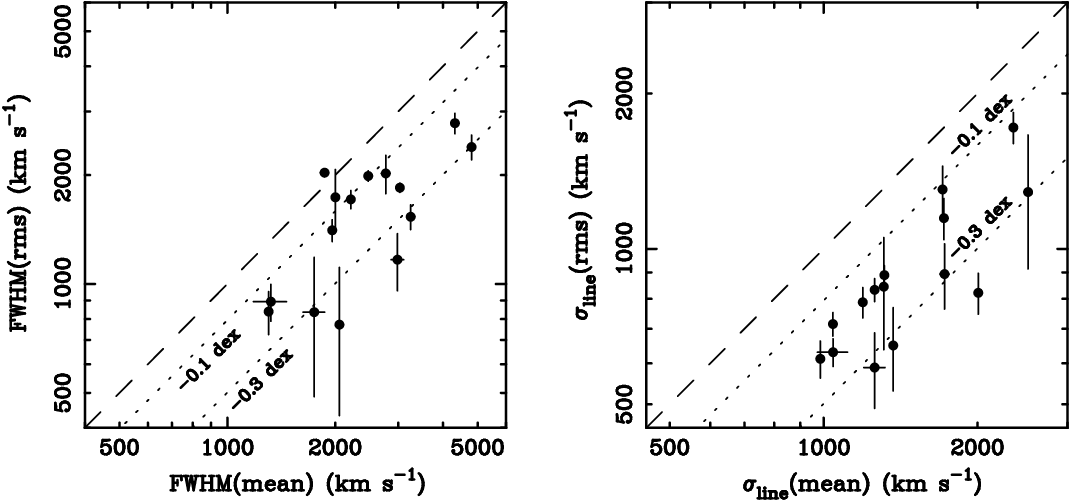}
  \caption{
  Comparisons of the FWHMs (left panel) and \sline\ (right panel) in mean and
  rms spectra. The two dotted lines in each panel mark where the widths in the
  rms spectra are 0.1 dex and 0.3 dex lower, respectively.
  }
  \label{fig-width}
\end{figure*}

Figure \ref{fig-width} shows the comparisons between the mean and rms spectra
for the FWHM (left panel) and the \sline\ (right panel). For most objects in
our sample, the \hb\ emission line in the rms spectrum is narrower than that
in the mean. For both FWHM and \sline, the values in the rms spectra are
$\sim$0.1 dex to $\sim$0.3 dex narrower than the values in the mean spectra.
Judging by their weak \oiii\ lines, the contamination of the narrow \hb\
component is negligible for most objects in our sample. Thus the narrower
broad line in the rms spectrum could be the natural consequence of
photoionization: the responsivity is higher at larger radius, and thus the
line core is more variable than the wings \citep[see,
e.g.,][]{korista04,goad14}.

\begin{figure}
  \centering
  \includegraphics[width=0.45\textwidth]{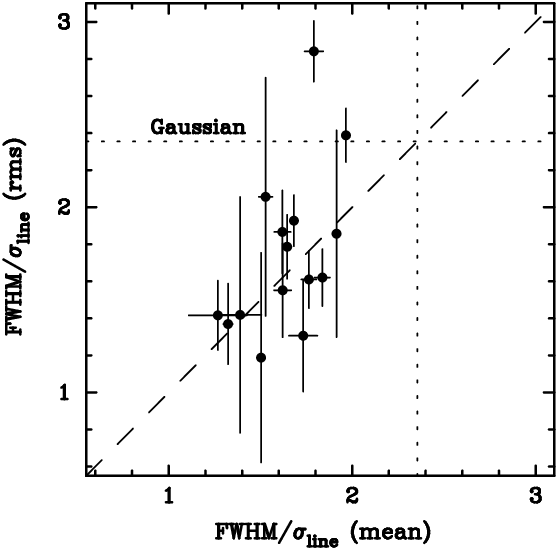}
  \caption{
  The FWHM/\sline\ ratio in mean and rms spectra. The dotted lines mark the
  value for a Gaussian.
  }
  \label{fig-widthratio}
\end{figure}

Figure \ref{fig-widthratio} shows the comparison of the FWHM/\sline\ ratio in
mean and rms spectra. The value of this ratio in the rms spectrum and
that in the mean spectrum are roughly equal for most objects, and both are
smaller than that of a Gaussian (dotted lines). Compared to the distribution
in \citet{peterson04} (their Figure 9), our sample is biased to have more
Lorenzian-like profiles of \hb\ in the rms spectra, which may be a feature of
high-accreting AGNs \citep[e.g.,][]{veron-cetty01}.

The dimensionless factor $f$ incorporates all the unknown effects including,
e.g., the geometry, kinematics, and inclination of the BLR, and therefore it
should not necessarily be the same for different objects. It is possible to
obtain the value of $f$ for individual object by the direct modeling method
\citep{pancoast11} if the data quality allows. However, in practice, the
factor $f$ is often estimated as an average of a sample obtained through a
calibration that uses the masses determined by other methods, e.g., the
$M_{\rm BH}$--$\sigma_\ast$ relation \citep[e.g.,][]{onken04,grier13}.
\citet{ho14} found that the uncertainty in $f$ can be reduced by considering
the bulge classification of the host galaxy.

The column (6) of Table \ref{tab-mass} lists the virial products defined as $c
\tau \sigma_{\rm line}^2 / G$, where the \sline\ in the rms spectrum is adopted
following \citet{peterson04}. The value of 4.31 estimated by \citet{grier13}
is adopted as an averaged $f$ here, and the results of \mbh\ are reported in
column (7). The optical luminosities for the host galaxy ($\lambda L_{\rm
\lambda,gal}$) and the AGN ($\lambda L_{\rm \lambda,AGN}$) at 5100 \AA\ are
also given as columns (8) and (9), respectively, derived from the fluxes of
corresponding components obtained in the best fit to the mean spectrum with
cosmological parameters of $H_0=72~{\rm km~s^{-1}}$ Mpc$^{-1}$,
$\Omega_{m}=0.3$, and $\Omega_\Lambda=0.7$. The errors of $\lambda L_{\rm
\lambda,AGN}$ were derived from the standard deviations in the \fc, thus
represent the variability of the AGN continuum during our observations. The
Galactic extinction has been corrected using an extinction law assuming $R_V$
= 3.1 \citep{odonnell94} and the $V$-band extinction magnitude listed in Table
\ref{tab-obs}. Note that host galaxy contribution from spectral decomposition
given here is uncertain due to the probable mismatch between the simple
stellar population template we used and the actual host starlight spectrum.
The host galaxy fraction from the photometry decomposition is more reliable
and preferred \citep[e.g.,][]{rakshit19}. The analysis of \textit{HST} images
to determine accurate host galaxy contamination and bulge properties (thus
enabling a better $f$ estimation) will be performed in a future contribution,
leading to improved optical luminosity measurement and \mbh\ estimation.

\section{Comparison with Previous Time Lag Measurements}
\label{sec-cmp}

\begin{figure*}
  \centering
  \includegraphics[width=0.9\textwidth]{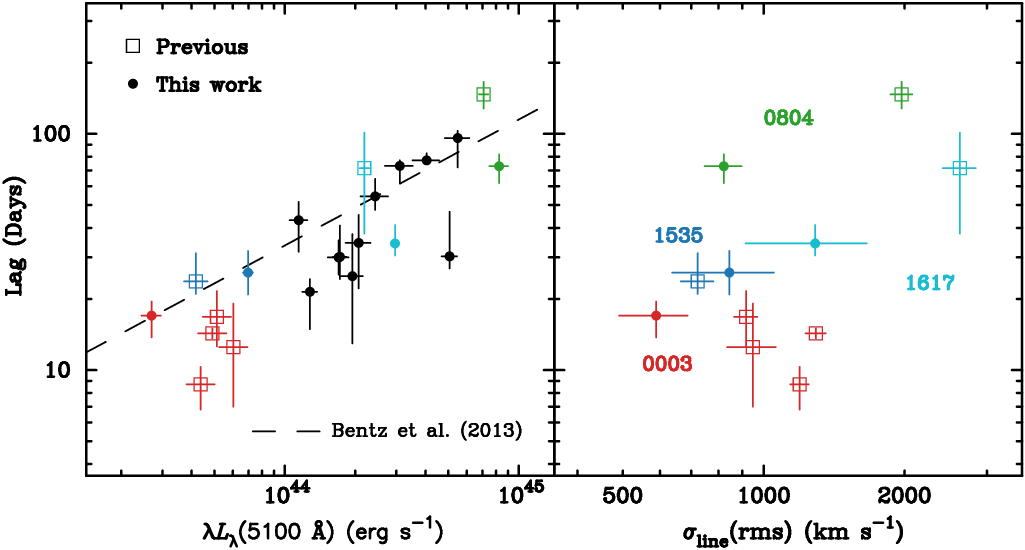}
  \caption{
  Time lag versus the optical luminosity (left) and line width (right), for
  the four objects with previous reverberation measurements (different colors
  for different objects). The results in this work are plotted as solid
  circles, while previous measurements are plotted as open squares. The 11
  newly detected objects are also plotted in black for comparison. The dashed
  line is the $R_{\rm BLR}$--$L$ relation in \citet{bentz13}.
  }
  \label{fig-lagcmp}
\end{figure*}

Four objects in our sample had \hb\ time lag measurements determined prior to
our campaign. The multiple reverberation mapping studies of NGC 5548 show that
for this single object, the \hb\ time lag is correlated to the optical
luminosity, as the BLR radius--luminosity relation in a sample
\citep[e.g.][]{bentz13,cho20}, but with a steeper slope \citep{lu16}. The left
panel of Figure \ref{fig-lagcmp} shows a comparison between the \hb\ time lag
and the optical luminosity for our sample. The four objects that have previous
reverberation measurements are plotted in different colors. The results from
previous studies are plotted as open squares, while the measurements in this
work are displayed as solid circles. Other objects in our sample with \hb\
time lag measured for the first time are also included in this diagram (in
black). For comparison, the $R_{\rm BLR}$--$L$ relation of \citet{bentz13} is
plotted as the dashed line. About half of the objects in our sample have \hb\
time lags consistent with the values predicted by the relation, while some
could have their measured \hb\ time lags nearly 0.3 dex shorter. Detailed
analysis of the $R_{\rm BLR}$--$L$ relation of the current sample, including
the dependence of the deviation to the relation on other properties, will be
presented in a future contribution with more accurate determination of the
host galaxy contamination by analyzing the \textit{HST} images.

While the \hb\ lag and the optical luminosity for the sample show a strong
positive correlation, the relationship between the two quantities for the four
objects with multiple measurements shows diverse behavior. On the other
hand, the time lag and the width of a single line have been observed to be
correlated in several objects with multiple reverberation studies
\citep{peterson04}, indicating that the clouds in their BLRs are well
virialized. The right panel of Figure \ref{fig-lagcmp} plots the \hb\ time
lag versus \sline(rms) in this work and previous studies for the four objects
with multiple measurements. Combining these two figures, we summarize the
changes in the measured time lags of these four objects individually as
follows.

\textit{PG 0003+199 (Mrk 335).} There are four previous reverberation mapping
measurements of PG 0003+199 by \citet{kassebaum97}, \citet{peterson98a},
\citet{grier12}, and \citet{hu15}. The median sampling intervals in
\citet{kassebaum97} and \citet{peterson98a} are roughly 7 days, while the
observations in \citet{grier12} and \citet{hu15} have much higher cadences of
nearly one day. The \hb\ time lags and optical luminosities at 5100 \AA\ are
plotted in the left panel of Figure \ref{fig-lagcmp} as red open squares. The
optical luminosity in this work is the lowest even without the host galaxy
contamination correction, while the time lag is the longest (red solid
circle). Thus, the responsivity weighted radius and the optical luminosity do
not follow the positive correlations observed in reverberation mapping sample
and in some other well-known AGN (e.g., NGC 5548).

On the other hand, the observed \hb\ time lag appears inversely proportional
to the square of the line width characterized as line dispersion in the rms
spectra (the right panel of Figure \ref{fig-lagcmp}, red data points),
especially for the most recent two observations (\citealt{hu15} and this work;
the former measured a lag of $8.7_{-1.9}^{+1.6}$ days and a \sline\ of
$1194\pm54$ \kms). Our observations produced the longest time lag and the
narrowest line width measurements, suggesting that the \hb-emitting clouds are
still virialized.

\textit{PG 0804+761.} The \hb\ time lag of PG 0804+761 in the observations by
\citet{kaspi00} is $146.9_{-18.9}^{+18.8}$ days (from the reanalysis of
\citealt{peterson04}), which is about two times as long as that measured in
this work. The luminosities of this object during the two campaigns are
actually very close (see the green data points in Figure \ref{fig-lagcmp}).
The FWHM(rms) of the two campaigns are also similar, according to the
determination by \citet{peterson04}, but \sline(rms) is two times larger
there. The time sampling rate in \citet{kaspi00} (70 data points in $\sim$7
years) is much lower than in this work (149 data points in $\sim$3 years). The
time lag can be overestimated for undersampling (see the discussion in
\citealt{grier08}), or the longer lag could be caused by some secular trends
in the \hb\ light curve. We have found similar much shorter time lags
compared to those of \citet{kaspi00} in our high cadence data for PG 2130+099
and PG 0026+129 \citep{hu20b, hu20a}.

\textit{PG 1535+547 (Mrk 486).} There is one previous reverberation mapping
measurement of PG 1535+547 by \citet{hu15}, who obtained a \hb\ time lag of
$23.7_{-2.7}^{+7.5}$ days. The two time lags are equivalent considering the
uncertainties. As shown in Figure \ref{fig-lagcmp} (blue data points), the
optical luminosity determined in this work is much higher than that in
\citet{hu15}, while the line widths are also similar considering the
uncertainties in the measurements (723$\pm$58 \kms\ for \sline\ in the rms
spectrum there). As in the case of PG 0003+199 above, the
responsivity-weighted radius of the BLR (observed time lag) is not correlated
with the optical luminosity, but the line-emitting clouds are still well
virialized.

\textit{PG 1617+175 (Mrk 877).} PG 1617+175 was observed by \citet{kaspi00}
and reanalyzed by \citet{peterson04}, which yielded an \hb\ time lag of
$71.5_{-33.7}^{+29.6}$ days that is also almost two times as high as the
measurement here. As in the case of PG 0804+761 above, such a large difference
is not likely to be related to the changes in luminosity or line width,
but could be caused either by the undersampling in \citet{kaspi00} (only 35
epochs in $\sim$7 years), or an unknown long-term evolution of the emission
line.

In summary, the effects of the time-sampling properties have to be considered
when comparing the measured time lags of a single object between different
reverberation campaigns. The short \hb\ time lags in our measurements of PG
0804+761 and PG 1617+175 that are only half as large as those in
previous observations with sparse sampling, as well as the cases of PG
2130+099 and PG 0026+129 in \citet{hu20b} and \citet{hu20a}, indicate
that high quality of both sampling density and duration are important for
measuring time lags as long as several tens to a hundred days. The previous
time lag measurements at the high-luminosity end of the $R_{\rm BLR}$--$L$
relation could be overestimated by a factor of two due to undersampling. The
sample in this work will be used to investigate this issue after correcting
for the host-galaxy contamination in a future contribution.

For PG 0003+199 and PG 1535+547, the comparison between our measurements and
previous results that also had high sampling rates shows that time lag is
inversely correlated with the line width, consistent with $\tau \propto
v^{-2}$ as a consequence of virialization that the motions of the
line-emitting gas are in equilibrium and dominated by the gravity of the
central supermassive black hole. However, the responsivity-weighted BLR radius
is not well correlated with the luminosity, as indicated by the
photoionization calculation if the inner and outer radii of the BLR are not
changed \citep[e.g.,][]{netzer20}. 

\section{Summary}

We started a long-term reverberation mapping campaign in May 2017, aimed
to spectroscopically monitor a large sample of PG quasars with both
long duration and high cadence for reliable time lag measurements. In this
paper, we provide the results of the observations for a sample of 15 PG
quasars with redshift up to 0.17 during the first three years of this campaign
with median sampling intervals of 3--9 days, as summarized below.

\begin{enumerate}
  \item Significant reverberations between the broad \hb\ emission line and
    the AGN continuum are observed, and reliable time lags are measured for 15
    PG quasars. The rest-frame time lags range from $17.0_{-3.2}^{+2.5}$ days
    in PG 0003+199 to $95.9_{-23.9}^{+7.1}$ days in PG 1402+261. In 11 of
    these objects this is measured for the first time.
  \item The \hb\ time lags of PG 0804+761 and PG 1617+175 are only half as
    large as those given by previous campaigns with sparse sampling.
    Considering that such differences in time lags can not be interpreted
    by changes in neither luminosities nor line widths, it is more probable
    that the time lags were overestimated in previous campaigns due to
    undersampling.
  \item For each of PG 0003+199 (Mrk 335) and PG 1535+547 (Mrk 486), a
    comparison with previous high-cadence studies indicates that the \hb\ time
    lag does not correlate with the optical luminosity, but depends on the
    line width, as virialization requires.
  \item The widths of the \hb\ emission line, characterized by either FWHM or
    line dispersion, are systematically narrower in the rms spectra than those
    in the mean spectra. The profiles of both mean and rms spectra are more
    Lorenzian-like.
  \item The masses of the central black holes are estimated for the sample.
    The values range between $0.50_{-0.19}^{+0.18}$ (PG 0003+199) and
    $19.17_{-2.73}^{+2.98}$ (PG 1626+554) in units of $10^7 M_\odot$.
\end{enumerate}

More analysis to this data set, e.g., the time lags of other emission lines,
host galaxy properties, and velocity-resolved delays, will be performed in
future contributions.

\acknowledgments
We are grateful to the referee for suggestions that helped to improve the
paper. We acknowledge the support of the staff of the CAHA 2.2m telescope.
This work is based on observations collected at the Centro Astron\'omico
Hispanoen Andaluc\'ia (CAHA) at Calar Alto, operated jointly by the Andalusian
Universities and the Instituto de Astrof\'isica de Andaluc\'ia (CSIC). This
research is supported by the National Key R\&D Program of China
(2016YFA0400701, 2016YFA0400702), by the National Science Foundation of China
(11721303, 11773029, 11833008, 11873048, 11922304, 11973029, 11991051,
11991052, 11991054, 12003036, 12022301), by the Key Research Program of
Frontier Sciences of the Chinese Academy of Sciences (CAS; QYZDJ-SSW-SLH007),
by the CAS Key Research Program (KJZDEW-M06), and by the Strategic Priority
Research Program of the CAS (XDB23000000, XDB23010400). MB enjoyed support
from the Chinese Academy of Sciences Presidents International Fellowship
Initiative, Grant No.2018VMA0005.  JA acknowledges financial support from the
State Agency for Research of the Spanish MCIU through the ``Center of
Excellence Severo Ochoa'' award to the Instituto de Astrof\'isica de
Andaluc\'ia (SEV-2017-0709).

\appendix

\section{Comparing Integration and Fitting}
\label{sec-methodcmp}

For light-curve measurements, spectral fitting has the advantage of
decomposing the contamination due to \heii\ \citep[e.g.,][]{hu20a} and the
host galaxy \citep[e.g.,][]{hu16}. On the other hand, the limited S/N and
spectral resolution of the individual-epoch spectrum, and the oversimplified
and probably slightly mismatched template of the host starlight, introduce
uncertainties sometimes even larger than those produced by simple integration,
as demonstrated in the case of PG 2130+099 in \citet{hu20b}. To compare the
uncertainties in the light-curve measurements by integration and fitting,
input-output simulations were performed under conditions representative of the
spectral quality recorded in this study and expected host galaxy fractions.

\begin{figure}
  \centering
  \includegraphics[width=0.475\textwidth]{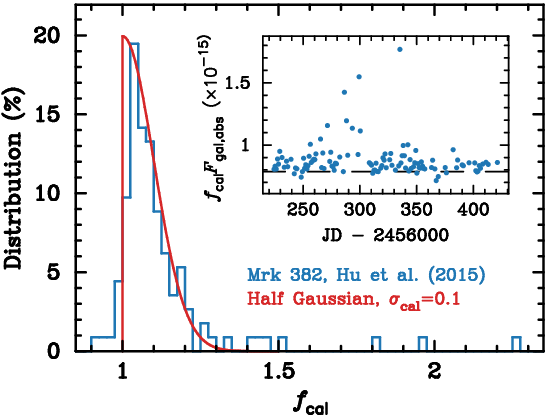}
  \caption{
  The distribution of the apparent-flux-variations factor \fcal\ for Mrk 382
  in \citet{hu15} (blue histogram), which can be described by a half Gaussian
  with \sigcal\ = 0.1 (red curve). The inserted plot shows the original light
  curve of the observed host galaxy flux \fgal\ = \fcal\fgalabs\ in
  \citet{hu15}. The horizontal dashed line marks the value of \fgalabs\
  adopted.
  }
  \label{fig-afv}
\end{figure}

Because of the different slit losses for the host galaxy (an extended source)
and the comparison star (a point source), the recorded flux of the host galaxy
\fgal\ in our observations shows apparent flux variations: \fgal\ =
\fcal\fgalabs, where \fgalabs\ is the non-varying absolute flux of the host
galaxy, and \fcal\ is a factor accounting for the deviations introduced by
seeing effects and inexact centering. (see Appendix A of \citealt{hu15} for a
detailed explanation, and also Appendix A of \citealt{lu19}). For Mrk 382 in
\citet{hu15}, the observed \fgal\ light curve (the inserted plot of Figure
\ref{fig-afv}; see also Figure 9 of \citealt{hu15}) shows that \fcal\ (the
blue histogram in Figure \ref{fig-afv}) can be roughly described as a
half-normal distribution peaked at 1 and \sigcal\ = 0.1 (the red curve). Poor
seeing and mis-centering always lead to relatively more slit loss for the
comparison star than the host galaxy, and hence \fcal\ is larger than unity.
The value of \sigcal\ can be affected by the size of the host galaxy, the
width of the slit, and the variability of the seeing. Thus, in the simulations
below, two values of \sigcal\ (0.1 and 0.2) were adopted, along with the case
of no apparent flux variations (\sigcal\ = 0).

The simulations are similar to those done in \citet{hu08b}. Simulated spectra
were generated including following components: (1) a single power law with a
spectral index of $-1.5$ and a flux of unity at 5100 \AA\ (\fagn), (2) a host
galaxy component determined by a template with an age of 11 Gyr and solar
metallicity from \citet{bruzual03}, (3) \feii\ emission (EW = 60 \AA, FWHM =
2000 \kms), (4) a Gauss-Hermite profile (EW = 80 \AA, FWHM = 2500 \kms, $h_3$
= 0, $h_4$ = 0.12) for the broad \hb\ emission line, (5) narrow lines
including \hb\ (EW = 4 \AA, FWHM = 1200 \kms) and \oiii\
$\lambda\lambda$4959,5007 (EW = 40 \AA\ for $\lambda$5007 line). These
parameters were chosen to be similar to those in our observed spectra of PG
0003+199. The strength of the host galaxy component was determined by the flux
ratio of \fgalabs/\fagn\ and \fcal, which was generated randomly according to
a half-normal distribution with \sigcal. The simulated spectra were set to
have the same dispersion of 4.47 \AA\ as the observed spectra in this work,
and Gaussian random noise was added to yield a S/N of 50 pixel$^{-1}$, which
is typical for our observed individual-epoch spectra. For each pair of values
for \fgalabs/\fagn\ and \sigcal, 100 simulations were generated. The fluxes of
the 5100 \AA\ continuum and the broad \hb\ were then measured by integration
and spectral fitting, respectively. The standard deviations in the continuum
(\sigc) and \hb\ fluxes (\sighb) measured in the 100 simulations can be
considered as the uncertainties given by each method.

\begin{figure}
  \centering
  \includegraphics[width=0.475\textwidth]{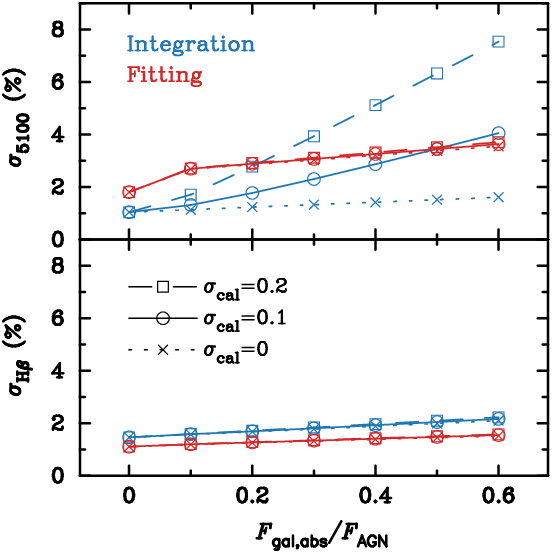}
  \caption{
  Input-output simulations of the uncertainties in the continuum (top panel)
  and \hb\ (bottom panel) flux measurements obtained by integration (blue)
  and fitting (red). The relative strength of the host galaxy with respect to
  the AGN, \fgalabs/\fagn, was adjusted between 0 and 0.6. Three values of the
  apparent-flux-variations strength, \sigcal, were simulated.
  }
  \label{fig-methodcmp}
\end{figure}

Figure \ref{fig-methodcmp} shows the \sigc\ (top) and the \sighb\ (bottom)
determined by integration (blue) and fitting (red) from simulations with
different combinations of \fgalabs/\fagn\ and \sigcal. Several conclusions are
indicated by these simulations. (1) The uncertainties in \hb\ flux
measurements by the two methods are comparable and weakly depend on the
strength of the host galaxy. This has been demonstrated in the case of Mrk 382
(\fgalabs/\fagn\ $\approx$ 0.8) where the integrated \hb\ light curve (Figure
2 in \citealt{wang14}) yielded a time lag consistent with that given by
fitting in \citet{hu15}. (2) The uncertainty in the continuum flux by fitting
does not depend on \sigcal, but increases slowly from $\sim$2\% to $\sim$3\%
with increasing host contribution. (3) The uncertainty in the continuum flux
by integration is affected by \sigcal, as a consequence that the integrated
5100 \AA\ flux \fc\ = \fagn\ + \fcal\fgalabs. (4) Integration is better than
fitting for measuring the continuum flux when \fgalabs/\fagn\ $<$ 0.5 if
\sigcal\ $\leq$ 0.1, or \fgalabs/\fagn\ $<$ 0.2 if \sigcal\ $\leq$ 0.2. This
is the situation in the case of PG 2130+099 in \citet{hu20b}, which has a
\fgalabs/\fagn\ $\approx$ 0.3.

From the best fits to the mean spectra, all the 15 objects in this work have
flux ratios of the host galaxy to the AGN continuum at 5100 \AA\ $\lesssim$
0.3 except PG 0838+770 (whose ratio is 0.55) (See Table \ref{tab-mass}). In
addition, none of these objects has a broad \heii\ emission line so strong
that it distorts the integration, as happened in the case of PG 0026+129 in
\citet{hu20a}. Thus, integration is adopted for light-curve measurements in
this work. The results of spectral line fitting on \hb\ will be presented for
comparison in a future contribution on the time lags of \feii\ emission.

\end{document}